\documentclass{article}
\usepackage{bm,latexsym,amsmath,amssymb,amsfonts,fancyhdr,color,multirow,slashed,cite}
\usepackage{graphicx}
\usepackage{lmodern}
\usepackage[a4paper,bottom=3cm,top=2.5cm,head=0mm,width=17cm,dvipdfm]{geometry}
\usepackage[usenames,dvipsnames,svgnames,table]{xcolor}
\usepackage[colorlinks=true,
            linkcolor=blue,
            urlcolor=blue,
            citecolor=JungleGreen,          
        bookmarks=true,
        bookmarksnumbered=true,
        breaklinks=true,
        pdfstartview=FitBH]{hyperref}
\usepackage[dotinlabels]{titletoc}
\usepackage{titlesec}
\usepackage{authblk,ulem}
\usepackage{float}
\usepackage{multirow}
\newcommand{\stkout}[1]{\ifmmode\text{\sout{\ensuremath{#1}}}\else\sout{#1}\fi}

\numberwithin{equation}{section}
\allowdisplaybreaks[4]

\titlelabel{\thetitle.\quad \hspace{-0.8em}}
\titlecontents{section}
              [1.5em]
              {\vspace{4mm} \large \bf}
              {\contentslabel{1em}}
              {\hspace*{-1em}}
              {\titlerule*[.5pc]{.}\contentspage}
\titlecontents{subsection}
              [3.5em]
              {\vspace{2mm}}
              {\contentslabel{1.8em}}
              {\hspace*{.3em}}
              {\titlerule*[.5pc]{.}\contentspage}
\titlecontents{subsubsection}
              [5.5em]
              {\vspace{2mm}}
              {\contentslabel{2.5em}}
              {\hspace*{.3em}}
              {\titlerule*[.5pc]{.}\contentspage}

\newcommand{\titledef}
{
CP-violating Higgs Di-tau Decays: Baryogenesis and Higgs Factories
}

\hypersetup{ pdfauthor = {Shao-Feng Ge},
         pdftitle = {\titledef}, 
         pdfsubject = {}, 
             pdfkeywords = {}, 
         pdfcreator = {LaTeX with hyperref package},
         pdfproducer = {dvips + ps2pdf} 
         }

\definecolor{gesfpurple}{rgb}{0.47,0.19,0.42}

\definecolor{gesflanse}{rgb}{0.00,0.50,0.50}

\definecolor{gesfblue}{rgb}{0.08,0.42,0.76}

\definecolor{gesfred}{rgb}{1,0,0}

\definecolor{gesfwhite}{rgb}{1,1,1}

\definecolor{gesfblack}{rgb}{0,0,0}

\definecolor{Orange}{cmyk}{0,0.61,0.87,0}
\definecolor{JungleGreen}{cmyk}{0.99,0,0.52,0}
\definecolor{OliveGreen}{cmyk}{0.64,0,0.95,0.40}
\definecolor{Brown}{cmyk}{0,0.81,1,0.60}
\definecolor{RoyalBlue}{cmyk}{0.71,0.53,0,0.12}
\definecolor{Gray}{cmyk}{0,0,0,0.40}
\definecolor{LightPink}{cmyk}{0.0,0.25,0,0}
\definecolor{LLightPink}{cmyk}{0.0,0.10,0,0}
\definecolor{LightBlue}{cmyk}{0.25,0,0,0}
\definecolor{LightGray}{cmyk}{0,0,0,0.2}

\newcommand{\gsec}[1]{{\hypersetup{linkcolor=red}Sec.~\ref{#1}\hypersetup{linkcolor=blue}}}

\newcommand{\geqn}[1]{\hypersetup{linkcolor=blue}Eq.~(\ref{#1})\hypersetup{linkcolor=blue}}
\newcommand{\gfig}[1]{{\hypersetup{linkcolor=violet}Fig.~\ref{#1}\hypersetup{linkcolor=blue}}}
\newcommand{\gtab}[1]{{\hypersetup{linkcolor=gesflanse}Table~\ref{#1}\hypersetup{linkcolor=blue}}}

\newcommand{\gang}[1]{{\color{blue}  #1}}

\begin{document}
\fontsize{12pt}{14pt}\selectfont

\title{\begin{flushright}
       \mbox{\normalsize ACFI-T20-18}
       \end{flushright}
             \medskip
       \textbf{\fontsize{14pt}{16pt}\selectfont \titledef}} 
\author[1,2]{{\large Shao-Feng Ge}\footnote{\href{mailto:gesf@sjtu.edu.cn}{gesf@sjtu.edu.cn}}\,\,}
\author[3]{{\large Gang Li}\footnote{\href{mailto:ligang@umass.edu}{ligang@umass.edu}}
}
\author[1]{{\large Pedro Pasquini}\footnote{\href{mailto:ppasquini@sjtu.edu.cn}{ppasquini@sjtu.edu.cn}}
}
\author[1,2,3,4]{{\large Michael J. Ramsey-Musolf}\footnote{\href{mailto:mjrm@sjtu.edu.cn}{mjrm@sjtu.edu.cn}}
}
\affil[1]{Tsung-Dao Lee Institute \& School of Physics and Astronomy, Shanghai Jiao Tong University, Shanghai 200240, China}
\affil[2]{Shanghai Key Laboratory for Particle Physics and Cosmology, Shanghai Jiao Tong University, Shanghai 200240, China}
\affil[3]{Amherst Center for Fundamental Interactions, Department of Physics, University of Massachusetts, Amherst, MA 01003, USA.}
\affil[4]{Kellogg Radiation Laboratory, California Institute of Technology, Pasadena, CA 91125, USA.}
\date{\today}

\maketitle

\begin{abstract}
\fontsize{12pt}{14pt}\selectfont
We demonstrate how probes of CP-violating observables in Higgs di-tau decays at prospective future lepton colliders could provide a test of weak scale baryogenesis with significant discovery potential. 
Measurements at the Circular Electron Positron Collider, for example, could exclude a CP phase larger than $2.9^\circ$ ($5.6^\circ$) at 68\% (95\%) C.L. assuming the Standard Model value for magnitude of the tau lepton Yukawa coupling. 
Conversely, this sensitivity
would allow for a $5\,\sigma$ discovery for 82\% of the CP phase range
$[0,2\pi)$. The reaches of the Future Circular Collider - ee and International Linear Collider are comparable. As a consequence, 
future lepton colliders could establish the presence of CP violation required by lepton flavored electroweak baryogenesis with 
at least $3\,\sigma$ sensitivity.
Our results illustrate that Higgs factories are not just
precision machines but can also make $\mathcal O(1)$
measurement of the new physics beyond the Standard Model.
\end{abstract}

\newpage

\section{Introduction}

The discovery of the Higgs boson at the Large Hadron Collider (LHC) \cite{Aad:2012tfa,Chatrchyan:2012ufa} and subsequent measurements of its properties strongly favor the mechanism of electroweak symmetry-breaking (EWSB) given by the Standard Model (SM) of particle physics.
In particular, the SM predicts each Higgs boson-fermion Yukawa coupling to be purely real, with magnitude proportional to the fermon mass.
However, LHC measurements have confirmed this prediction
up to only $\mathcal O(10)\%$ precision \cite{Chatrchyan:2014nva,Aad:2015vsa,Aaboud:2018pen, CMS:2020dvp}
, leaving considerable 
room for physics beyond the Standard Model (BSM) in Higgs boson-fermion interactions. Precision Higgs boson studies aim to explore these BSM possibilities.
Proposed future
lepton colliders, including the Circular Electron-Positron Collider (CEPC)~\cite{CEPCStudyGroup:2018ghi}, Future Circular Collider (FCC)-ee \cite{Abada:2019zxq},
and International Linear Collider (ILC) \cite{Baer:2013cma}, are designed for this purpose.

The motivations for BSM Higgs interactions are well known, including solutions to the hierarchy problem, generation of neutrino mass, dark matter, and the cosmic baryon asymmetry (BAU). In what follows, we focus on the possibility that the observation of CP-violating effects in Higgs-tau lepton interactions at a future lepton collider could provide new insight into the BAU problem. As pointed out by Sakharov\cite{Sakharov:1967dj}, a dynamical generation of the BAU requires three ingredients in the particle physics of the early universe: : (1) non-conserving baryon number (2) out-of-equilibrium dynamics (assuming CPT conservation) and (3) C and CP violation. While the SM contains the first ingredient in the form of electroweak sphalerons, it fails to provide the needed out-of-equilibrium conditions and requisite CP-violation. The presence of BSM physics in the dynamics of EWSB could remedy this situation -- the electroweak baryogenesis (EWBG) scenario (for a recent review, see, Ref.~\cite{Morrissey:2012db}). While flavor-diagonal CP-violating (CPV) interactions relevant to EWBG are strongly constrained by limits on permanent electric dipole moments (EDMs) of the electron, neutron, and neutral atoms~\cite{Engel:2013lsa,Chupp:2017rkp}
, the landscape for flavor non-diagonal CPV is less restricted. Here, we show that searches for CPV effects in the Higgs di-tau decays at future lepton colliders could provide an interesting probe of \lq\lq flavored EWBG"~\cite{Guo:2016ixx,Chiang:2016vgf,deVries:2018tgs,Fuchs:2020uoc,Xie:2020wzn}  in the lepton sector.

In addition to being theoretically well-motivated in its own right, EWBG has the additional attraction of experimental testability. Modifications of the SM scalar sector necessary for the EWBG out-of-equilibrium conditions provide a rich array of signatures accessible at the LHC and prospective future colliders \cite{Ramsey-Musolf:2019lsf}. 
The signatures of CP violation could appear in either low- or high-energy experiments. Our present focus is on the possible modification of the 
the $\tau$ Yukawa coupling by a nonzero CP phase $\Delta$
as defined in \geqn{eq:interaction} below and the resulting impact on
Higgs decay into a pair of $\tau$ leptons. In this context, it has been known for some time that 
the $\Delta$  phase can be measured at colliders \cite{DellAquila:1988sbm, DellAquila:1988bko,Grzadkowski:1995rx,Bower:2002zx,Worek:2003zp,Desch:2003mw,Desch:2003rw,Berge:2008dr,Berge:2008wi,Berge:2011ij,Berge:2015nua,Ellis:2004fs,Berge:2013jra}. An $\mathcal O(1)$ modulation in the relevant differential distribution
allows very accurate determination of $\Delta$.
The measurements at LHC can achieve a precision around 
10$^\circ$\cite{Harnik:2013aja,Askew:2015mda,Hagiwara:2016zqz,Bhardwaj:2016lcu,Han:2016bvf,Chen:2017nxp}
at 95\% confidence level (C.L.) by using the full
data with 3~ab$^{-1}$ of integrated luminosity. 
Future lepton colliders \cite{deBlas:2019rxi} can further
improve the measurements with higher integrated luminosity,
optimized energy for Higgs production, and cleaner
environment. Indeed, it was shown that the projected
sensitivity for the ILC can reach $4.3^\circ$
\cite{Jeans:2018anq} and
$2.9^\circ$ for the CEPC \cite{Chen:2017bff}
at $1\,\sigma$ level.
To our knowledge, a detailed study connecting this
sensitivity to the BAU has yet to appear in the literature.

In this work we thus study the capability of the CEPC, FCC-ee 
and ILC in probing $\Delta$ and the resulting prospects of testing lepton flavored EWBG scenario. The projected
sensitivities at future lepton colliders are much better than
the current LHC results \cite{CMS:2020rpr}.
In \gsec{sec:CP-violation},
we first make detailed comparison of several observables
(the neutrino azimuthal angle difference
$\delta \phi_\nu$, the polarimeter $\delta \phi_r$,
the acoplanarity $\phi^*$, and the $\Theta$ variable)
to show that the polarimeter \cite{Grzadkowski:1995rx,Kuhn:1982di} is not just
the optimal choice for probing $\Delta$ but can also
apply universally
to both the $\tau \rightarrow \pi \nu$ and
$\tau \rightarrow \rho \nu$ decay channels.
Then we use a simplified smearing scheme to simulate 
the detector responses and use $\chi^2$ minimization
to find the physical solution of neutrino/tau
momentum in \gsec{sec:simulation_details}. Based
on these, we find that the future lepton colliders
can make $5\,\sigma$ discovery of a nonzero CP phase for
82\% of the allowed range.
With a combination of these channels, the
$1\,\sigma$ sensitivities can reach
$2.9^\circ, 3.2^\circ$, and $3.8^\circ$
at the CEPC, FCC-ee, and ILC, respectively.
Our result is better than the previous study
for the ILC \cite{Jeans:2018anq} and the same as
Ref.~\cite{Chen:2017bff} for the CEPC. Notice
that although the leptonic decay mode of $\tau$
is also considered in addition to the two meson
decay modes with more usable events,
a matrix element based observable is adopted
\cite{Chen:2017bff} instead of the polarimeter
$\delta \phi_r$ as we do here, leading to  accidentally the
same result as ours.
Finally, in \gsec{sec:new_phys_BAU},
we follow Ref.~\cite{Guo:2016ixx} in analyzing the
implications for lepton flavored EWBG scenario with
$3\,\sigma$ sensitivity for the presence of CP violation
at the CEPC, FCC-ee and ILC. We summarize our findings in \gsec{sec:conclude}.

\section{CP Phase and Azimuthal Angle Distributions}
\label{sec:CP-violation}

The SM predicts the Higgs couplings with other SM particles
to be proportional to their masses, including the $\tau$
lepton,
$-y_\tau/\sqrt{2} \bar{\tau}_L \tau_R h +\mathrm{h.c.} = - m_\tau/v \bar{\tau}\tau h$
, where $h$ denotes the SM-like Higgs boson, $v=246$~GeV is the vacuum expectation value (VEV)  and $m_\tau$ is the tau lepton mass. Although $y_\tau$ is in general complex,
its CP phase can be rotated away without leaving any
physical consequences. However, this is not
always true when going beyond the SM. Any deviation
from the SM prediction, for either the $\tau$ Yukawa
coupling magnitude or the CP phase,
indicates new physics. We first study the CP phase
measurement by explicitly comparing various definitions of
differential distributions in this section and
then the detector response behaviors in 
\gsec{sec:simulation_details}. The influence of the
Yukawa coupling magnitude deviation from the SM prediction
will be discussed in later parts of this paper.

The $\tau$ Yukawa coupling can be generally parametrized as 
\begin{eqnarray}
  \mathcal{L}_{h\tau\tau}
=
  -\kappa_\tau 
  \dfrac{m_\tau}{v}
  \bar{\tau} 
  (
    \cos\Delta 
  +
    i\gamma_5\sin\Delta
  )
  \tau h
\label{eq:interaction}
\end{eqnarray}
with 
$\kappa_\tau$ being real and positive by definition,
and $\Delta\in [0,2\pi)$ in general. We will consider $\Delta \in [0,\pi]$ since the CP measurement is insensitive to the multiplication of $\kappa_\tau$ by $-1$~\cite{Harnik:2013aja}. 
The SM prediction for $\tau$ Yukawa coupling
can be recovered with $\kappa_\tau \rightarrow 1$ and
$\Delta \rightarrow 0$. A non-zero value of the
CP phase $\Delta$ indicates CP violation in the $\tau$
Yukawa coupling and can be connected to baryogenesis in
the early Universe \cite{Guo:2016ixx}. 

Because of the P- and T- violating nature of 
the second term in Eq.~(\ref{eq:interaction}),
the spin correlation among the two $\tau$ leptons
from a Higgs decay is an especially interesting probe
for constraining its value~\cite{Worek:2003zp}. 
In practice, one cannot measure the $\tau$ lepton
directly but must rely on its decay products. 
The two most promising channels are the $\tau$ decay
into $\pi^{\pm}$ ($\tau^\pm \rightarrow \pi^\pm \nu_{\tau^\pm}$)
and into $\rho^{\pm}$
($\tau^\pm \rightarrow \rho^\pm \nu_{\tau^\pm} \rightarrow \pi^\pm \pi^0 \nu_{\tau^\pm}$ ) with $\nu_{\tau^-} (\nu_{\tau^+})$ being the neutrino (anti-neutrino) from the decay of $\tau^- (\tau^+)$.
These two channels contribute 10.82\%
and 25.49\% of a single $\tau$ decay branching
fractions \cite{Zyla:2020zbs}, respectively.

\subsection{Observables}

For each $\tau$ decay, one decay plane
can be formed by its decay products. The generic azimuthal angle $\phi$
difference between the two decay planes is then
a good observable for probing $\Delta$, which can be expressed as
\begin{equation}
  \frac 1 {\Gamma}
  \frac {d \Gamma} {d \delta \phi}
=
  \frac 1 {2\pi}
\left[
  1
+ A \cos(2\Delta - \delta \phi)
\right]\;,
\label{eq:dSigmadPhi}
\end{equation}
where the coefficient $A$ depends on the choice
of observable.
Note that only for $\Delta$ differing from integer
multiples of $\pi/2$, this distribution will contain
a term odd in $\delta\phi$.
There exist a variety of observables that afford access to $\Delta$, which appears as the azimuthal angle difference in two decay planes. Here we review several possibilities and discuss the rationale for our choice of one of them.

\begin{figure}[t!]
\centering
\includegraphics[scale = 0.35]{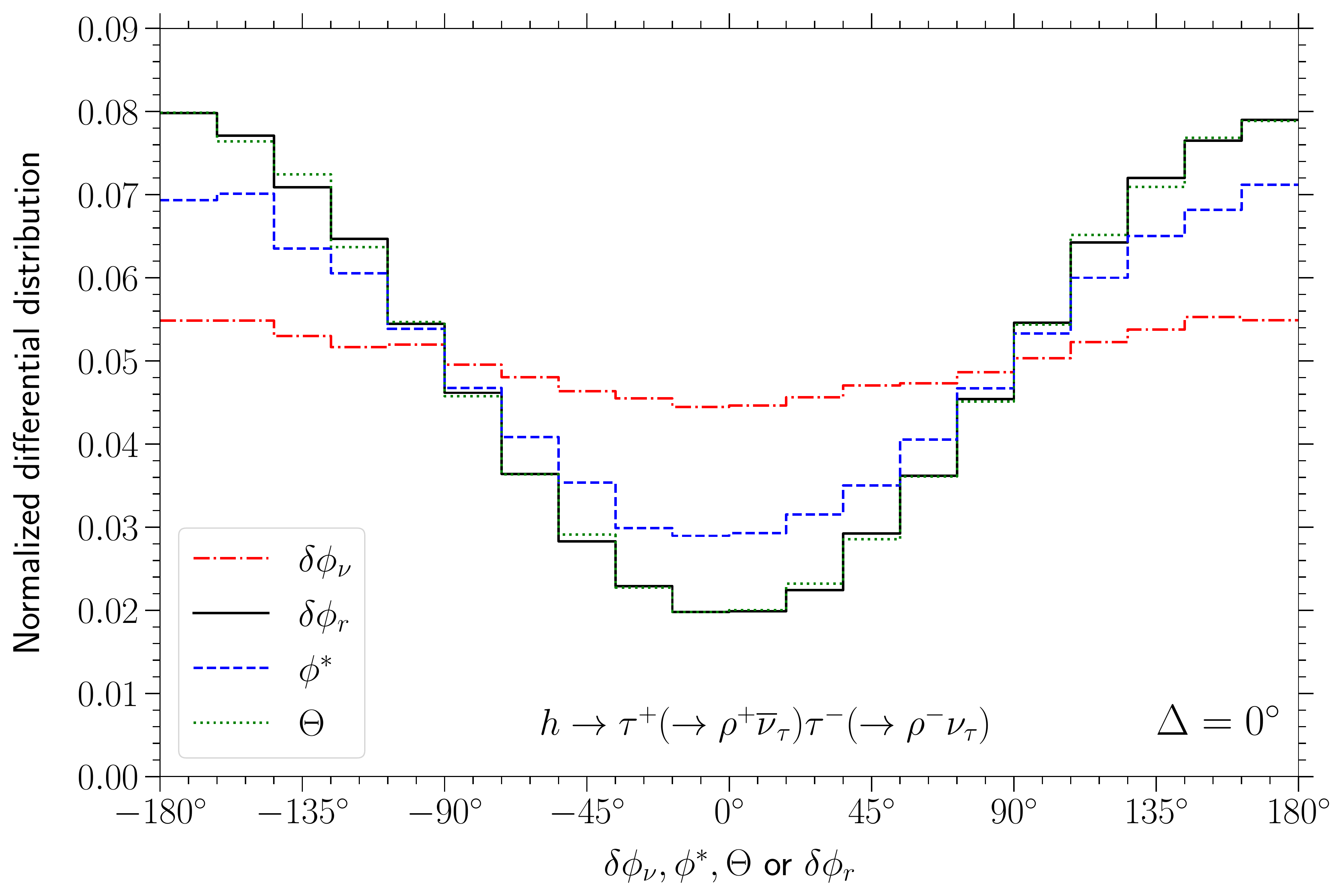}
\caption{The differential distributions of
$h\to \tau^+(\rightarrow \rho^+ \bar{\nu}_\tau) \tau^-(\rightarrow \rho^- \nu_\tau)$
for neutrino momentum $\delta \phi_\nu$
(red), polarimeter $\delta \phi_r$ (black),
acoplanarity $\phi^*$ (blue), and
the $\Theta$ variable (green) at the truth level for
$\Delta = 0^\circ$.}
\label{fig:comparison_vars}
\end{figure}

\noindent {\it Neutrino azimuthal angle difference.}
For both $\tau$ leptons decaying into a single
charged pion,
$\tau^\pm \rightarrow \pi^\pm \nu_{\tau^\pm}$,
the differential distribution of the neutrino
momentum azimuthal angle
difference \cite{Kuhn:1982di} is
\begin{equation}
  \frac 1 {\Gamma}
  \frac {d \Gamma (h \rightarrow \pi^+ \pi^- \nu_\tau \overline \nu_\tau)} {d \delta \phi_\nu}
=
  \frac 1 {2\pi}
\left[
  1
- \frac{\pi^2}{16} \cos(2\Delta - \delta \phi_\nu)
\right]\ \ \ ,
\end{equation}
where $\delta \phi_\nu \equiv \phi_\nu - \phi_{\overline \nu}$ and $\phi_\nu (\phi_{\overline \nu})$ are defined in the $\tau^-(\tau^+)$ rest frame. 
On the other hand, if both $\tau$'s decay to rho mesons,
$\tau^\pm\to \rho^\pm(\rightarrow \pi^\pm + \pi^0)\nu_{\tau^\pm}$, 
the differential distribution of the neutrino
azimuthal angle difference $\delta \phi_\nu$ becomes,
\begin{equation}
   \frac{1}{\Gamma} 
   \frac{d\Gamma(h\rightarrow \rho^+ \rho^- \nu_\tau \overline \nu_\tau)}
   {d\delta \phi_\nu}
=
  \frac{1}{2\pi}
  \left[
    1
  -
    \frac{\pi^2}{16}
    \left(\frac{m_\tau^2 - 2m_\rho^2}{m_\tau^2 + 2m_\rho^2}\right)^2 
    \cos(2\Delta - \delta \phi_\nu)
  \right]
\label{eq;angular_distribution_rho_nu}
\end{equation}
with a non-negligible suppression factor,
$(m_\tau^2 - 2m_\rho^2)^2 / (m_\tau^2 + 2m_\rho^2)^2 \sim 0.2$.
As shown in \gfig{fig:comparison_vars},
this significantly reduces the sensitivity to the
CP phase $\Delta$. The neutrino azimuthal angle
difference $\delta \phi_\nu$ is a good observable
for the $\tau \rightarrow \pi \nu_\tau$ decay, but
not for the $\tau \rightarrow \rho \nu_\tau$ channel.

\noindent {\it Polarimeter.} Since the azimuthal angle difference is not
necessarily the optimized choice and multiple
definitions of azimuthal angle have been invented.
In Refs.~\cite{Kuhn:1982di,Grzadkowski:1995rx,Jeans:2018anq}, the
azimuthal angle difference between the 
{\it polarimeter} vectors
${\bf r}_\pm$ was studied. For the $\tau$ decays,
the polarimeter vectors are defined as
\begin{subequations}
\begin{eqnarray}
    \tau^\pm\rightarrow \pi^\pm \nu_{\tau^\pm} 
& : &
  {\bf r}_\pm 
\equiv
- \hat{{\bf p}}_{\nu_{\tau^\pm}},
\label{eq:Polarimeter-pi}
\\
    \tau^\pm\rightarrow \rho^\pm (\to\pi^\pm \pi^0) \nu_{\tau^\pm}
& : &  
  {\bf r}_\pm 
\equiv
- \frac{1}{N_\pm}
\left[
  \hat{{\bf p}}_{\nu_{\tau^\pm}}
+ \frac{2m_\tau}
   {m^2_\rho - 4m^2_\pi}
   \frac{E_{\pi^\pm} - E_{\pi^0_\pm}}{E_{\pi^\pm} + E_{\pi^0_\pm}}
   \left({\bf p}_{\pi^\pm} - {\bf p}_{\pi^0_\pm}\right)
\right],
\qquad
\end{eqnarray}
\label{eq:polarimeter}
\end{subequations}
where ${\bf r}_\pm$ is calculated in the corresponding
$\tau^\pm$ rest frame, $(E_{\pi^0_\pm}, {\bf p}_{\pi^0_\pm})$ is the $\pi^0$
momentum in the $\tau^\pm$ decay, and $N_\pm$ is a normalization factor
to ensure $|{\bf r}_\pm | = 1$. Then the differential
distribution in \geqn{eq:dSigmadPhi} becomes
\begin{equation}
   \frac{1}{\Gamma} 
   \frac{d\Gamma}
   {d\delta \phi_r}
=
  \frac{1}{2\pi}
\left[
  1
- \frac{\pi^2}{16} \cos(2\Delta - \delta \phi_r)
\right],
\label{eq:angular_distribution_polarimeter}
\end{equation}
for both decay channels including the mixed mode,
$h \rightarrow \rho^\pm \pi^\mp \nu_\tau \overline \nu_\tau$.
From the neutrino azimuthal angle difference
$\delta \phi_\nu$ in \geqn{eq;angular_distribution_rho_nu}
to the one of the polarimeter in
\geqn{eq:angular_distribution_polarimeter}, the
amplitude gets amplified by a factor of 5 which is
a significant improvement.

The azimuthal angle difference
$\delta \phi_r \equiv \phi_{{\bf r}_+} - \phi_{{\bf r}_-}$
is defined with respect to the $z$ direction,
${\bf z} \equiv {\hat{\bf p}}_{\tau^-}$. In the Higgs rest frame,
\begin{equation}
  \tan \delta \phi_r
=
  \frac{{\hat{\bf p}}_{\tau^-} \cdot ({\bf r}_+ \times {\bf r}_-)}
  {
    {\bf r}_- \cdot {\bf r}_+ 
  - 
    ({\bf r}_+ \cdot {\hat{\bf p}}_{\tau^-})
    ({\bf r}_- \cdot {\hat{\bf p}}_{\tau^-}) 
  }.
\label{eq:azimuth_diff_definition}
\end{equation}
For the 
$\tau^\pm \rightarrow \pi^\pm \nu_{\tau^\pm}$ decay channel,
the polarimeter is along the neutrino momentum direction,
namely, ${\bf r}_\pm = - \hat{\bf p}_{\nu_{\tau^\pm}}$ as shown in \geqn{eq:Polarimeter-pi}.
When di-tau decay into pions it is the azimuthal angle difference $\phi_\nu$.
In contrast, the polarimeter
for the $\tau^\pm \rightarrow \rho^\pm \nu_{\tau^\pm}$ decay
channel does not coincide with any momentum of the
final-state particles. For illustration,
the distribution of $\delta\phi_r$ for $h\to \tau^+(\rightarrow \rho^+ \bar{\nu}_\tau) \tau^-(\rightarrow \rho^- \nu_\tau)$ is shown in \gfig{fig:comparison_vars}.

The 4-vector $r_\pm=(0,{\bf r_\pm})$ serves as the
effective spin of the corresponding $\tau^\pm$ leptons.
This becomes evident
in the total matrix element of the Higgs decay chain,
\begin{eqnarray}
  |\mathcal{M}^{\rm total}|^2
&\propto&
  {\rm Tr}
  \left[
    \left(\slashed p_{\tau^-} + m_\tau\right)
    \left(1 + \gamma_5 \slashed r_-\right)
    \mathcal{O}
    \left(\slashed p_{\tau^+} - m_\tau\right)
    \left(1 - \gamma_5 \slashed r_+\right)
    \overline{\mathcal{O}}
  \right],
\end{eqnarray}
with $\mathcal{O} \equiv \cos \Delta + i\gamma_5 \sin \Delta$, $\overline{\mathcal{O}} \equiv \gamma^0 \mathcal{O}^\dagger \gamma^0$  and $p_{\tau^\pm} \equiv (E_{\tau^\pm}, {\bf p}_{\tau^\pm})$ being the momentum of $\tau^\pm$.
If the Higgs boson decays to polarized $\tau$ leptons,
it should be the $\tau$ spin vector $s_\pm= (|{\bf p}_{\tau^\pm}|/m_\tau, E_{\tau^\pm}/m_\tau {\hat {\bf p}_{\tau^\pm}})$ that
appears in place of the polarimeter $r_\pm$. But since the
Higgs decay chain also contains contribution from the
$\tau$ decays, $s_\pm$ is replaced by $r_\pm$ to take the
extra effects into consideration. 

\noindent {\it Acoplanarity}. This observable was introduced in Ref.~\cite{Bower:2002zx,Worek:2003zp} for the
$\tau \rightarrow \rho \nu_\tau$ decay mode.
In the rest frame of the $\rho^+\rho^-$ system, the
$\rho$ momenta are back to back. The decay products of
$\rho^\pm$ form two decay planes and the angle difference
between them is defined as {\it acoplanarity} $\phi^*$,
\begin{equation}
  \tan \phi^*
\equiv
  \frac{{\hat{\bf p}}_{\rho^-} \cdot [({\bf p}_{\pi^+}\times {\bf p}_{\pi^0_+})\times ({\bf p}_{\pi^-}\times {\bf p}_{\pi^0_-})]}
  {
    ({\bf p}_{\pi^+}\times {\bf p}_{\pi^0_+})\cdot({\bf p}_{\pi^-}\times {\bf p}_{\pi^0_-})
  }\,.
\label{eq:azimuth_diff_definition_acopla}
\end{equation}
This interesting variable requires only
the knowledge of the directly observable momenta
of $\pi^\pm$ and $\pi^0$.
However, the oscillation amplitude
of the distribution is suppressed by around
30\% in comparison with the polarimeter
as shown in \gfig{fig:comparison_vars}.

\noindent {\it The $\Theta$ Variable.}
For $\tau \rightarrow \rho \nu_\tau$ decay,
a fourth observable similar to the usual acoplanarity angle, can be defined as,
\begin{equation}
  \tan \Theta
\equiv
  \frac{{\hat{\bf p}}_{\tau^+} \cdot ({\bf E}_+ \times {\bf E}_-)}
  {
    {\bf E}_- \cdot {\bf E}_+ 
  - 
    ({\bf E}_+ \cdot {\hat{\bf p}}_{\tau^+})
    ({\bf E}_- \cdot {\hat{\bf p}}_{\tau^+}) 
  },
\label{eq:azimuth_diff_definition_theta}
\end{equation}
with ${\bf E}_\pm$ taking analogy to the electromagnetic 
fields. In the $\tau^\pm$ rest frames, the ${\bf E}_\pm$
vector can be expressed as \cite{Harnik:2013aja},
\begin{align}
  {\bf E}_\pm 
& \equiv
  \frac{m^2_\rho - 4 m^2_\pi}{2 m_\tau}
\left[
  \frac {m^2_\tau - m^2_\rho}
        {m^2_\tau + m^2_\rho}
  \hat{{\bf p}}_{\nu_{\tau^\pm}}
+ \frac{2m_\tau}
   {m^2_\rho - 4m^2_\pi}
   \frac{(E_{\pi^\pm} - E_{\pi^0})}{(E_{\pi^\pm} + E_{\pi^0})}
   \left({\bf p}_{\pi^\pm} - {\bf p}_{\pi^0_\pm}\right)
\right].
\end{align}
Note that
\geqn{eq:azimuth_diff_definition_theta} is slightly more general than the one presented in Ref.~\cite{Harnik:2013aja}, where they take the approximation $({\bf E}_\pm \cdot {\hat{\bf p}}_{\tau^\pm})\approx 0$.
It is very interesting to see that \geqn{eq:azimuth_diff_definition_theta} has
very similar form as \geqn{eq:azimuth_diff_definition}
with the only difference of a proportional factor
$(m^2_\rho - 4 m^2_\pi) / 2 m_\tau$.
Due to these similarities,
the $\Theta$ variable has roughly the same sensitivity
as polarimeter, see
\gfig{fig:comparison_vars}.

The comparison in \gfig{fig:comparison_vars} shows that the
polarimeter $\delta \phi_r$ and the $\Theta$ variable
are the optimal ones. However, the $\Theta$ variable
needs both momenta of $\pi^\pm$ and $\pi^0$, limiting
its scope to only the $\tau \rightarrow \rho \nu_\tau$
decay mode. In contrast, the polarimeter method applies
for both channels by matching ${\bf r}_\pm$ with different
combination of final-state particle momenta as shown in \geqn{eq:polarimeter}. So we adopt the polarimeter
scheme in the following part of this paper.

\section{Measurements at Future Lepton Colliders}
\label{sec:simulation_details}

Future lepton colliders \cite{Benedikt:2020ejr} are
designed to produce millions of Higgs events. 
The three prominent candidate colliders
are the CEPC \cite{CEPCStudyGroup:2018ghi}, 
FCC-ee \cite{Abada:2019zxq} and ILC \cite{Behnke:2013lya}.
The CEPC experiment \cite{CEPCStudyGroup:2018ghi} is
expected to have around $1.1\times 10^6$ Higgs events.
This comes from an
integrated luminosity of 5.6\,ab$^{-1}$ with two interaction
points (IP) and 7 years of running at $\sqrt{s} = 240$ GeV.
The FCC-ee has a higher luminosity and 4 interaction points,
but runs in the Higgs factory mode for only 3 years resulting in a 5\,ab$^{-1}$ of integrated luminosity or equivalently
$1.0\times 10^6$ Higgs events \cite{Abada:2019zxq}.
The ILC, on the other hand, has a significantly lower
integrated luminosity at 2\,ab$^{-1}$,
but is able to produce polarized
electrons/positrons which increases the cross section
significantly, effectively raising its number of Higgs
production to $0.64\times 10^6$~\cite{Baer:2013cma}.
The configuration of these three experiments and
the expected numbers of Higgs events at the benchmark
luminosities have been summarized in 
\gtab{tab:accelerator_configuration} for comparison.
\begin{table}[H]
\centering
\begin{tabular}{r|ccc}
& Integrated luminosity  & $\sqrt s$ & Number of Higgs bosons \\
\hline
CEPC~\cite{CEPCStudyGroup:2018ghi}   &  5.6~ab$^{-1}$ & 240 GeV & $1.1\times10^{6}$ \\
FCC-ee~\cite{Abada:2019zxq}          &  5~ab$^{-1}$   & 240 GeV & $1.0\times10^{6}$ \\
ILC~\cite{Baer:2013cma}              &  2~ab$^{-1}$    & 250 GeV & $0.64\times10^{6}$ 
\end{tabular}
\caption{Configurations (integrated luminosity,
energy $\sqrt s$, and Higgs production rate) at the
future lepton colliders CEPC, FCC-ee, and ILC.}
\label{tab:accelerator_configuration}
\end{table}
  
In this section, we study the detector responses,
including the smearing effects, selection cuts,
and momentum reconstruction ambiguities. With around
$650 \sim 1100$ events, the uncertainty
at the level of $14\% \sim 18\%$ is much smaller than
the expected $60\%$ modulation in the CP measurement.
This allows a $5\,\sigma$ discovery potential for
approximately 80\% of the allowed range in $[0, 2 \pi)$ of the CP phase $\Delta$
and a determination of $\Delta$ with the accuracy of $2.9^\circ \sim 3.8^\circ$.

\subsection{Simulation and Detector Responses}

At lepton colliders, the Higgs boson is mainly
produced in the so-called Higgsstrahlung process,
$e^+ e^- \rightarrow Z h$, with an associated $Z$
boson. This channel allows a model-independent
measurement
of the Higgs properties thanks to the recoil mass
reconstruction method \cite{An:2018dwb}.
The Higgs event is first
selected by reconstructing the $Z$ boson
without assuming any Higgs coupling with the SM
particles. The Higgs boson momentum can be either
derived from the $Z$ boson momentum using energy-momentum
conservation or reconstructed from the Higgs decay
products. Since there are always two neutrinos in the
final state of $h \rightarrow \tau \tau$ events,
the $Z$ boson momentum is needed to reconstruct the
Higgs momentum as the initial condition of the Higgs
decay kinematics to fully recover the two neutrino
momenta.

We use MadGraph \cite{Alwall:2011uj} and TauDecay
\cite{Hagiwara:2012vz} packages to simulate
the spin correlation in the Higgs decay chains.
For a realistic simulation, both
detector response and statistical fluctuations have
to be taken into consideration.
In order to perform fast detector simulation we construct a simplified smearing algorithm which is validated by
comparing with Delphes~\cite{deFavereau:2013fsa} output.

Using the recoil mass method, the smearing should in principle be applied to the $Z$ momentum.
Nevertheless, since the Higgs and $Z$ bosons are back to back
in the center of mass frame, we can directly smear the Higgs momentum.
Defining the $z$-axis along the Higgs momentum, only its $P_z$
component is affected by the $Z$ boson decay modes while the
other two, $P_x$ and $P_y$, have independent smearing behaviors.
To select the Higgsstrahlung events, those with the
reconstructed $Z$ invariant mass outside the range
$80~{\rm GeV} < \sqrt{p^2_Z} < 100$ GeV are discarded.
The momentum uncertainties of Higgs smearing have been
summarized in the left part of
\gtab{tab:smearing_parameters}.

\begin{table}[ht]
    \centering
      \begin{tabular}{c} 
      Higgs Smearing\\
\begin{tabular}{c|c}
  Observables & Uncertainty       \\
\hline
  $P_{x,y}$ & 1.82 GeV      \\
  $P_z $ ($Z\rightarrow jj)$        &     2.3 GeV       \\
  $P_z $ ($Z\rightarrow l\overline l)$ &     0.57 GeV  
\end{tabular}
\\ 
\end{tabular}
   $\qquad$
    \begin{tabular}{c}
    Pion Smearing \\
    \begin{tabular}{c|c}
      Observables & Uncertainty \\
       \hline 
       $\phi$        &     0.0002$|\eta|$+ 0.000022  \\
       $\eta$        &     0.000016$|\eta|$ + 0.00000022  \\
       $|{\bf p}_T|$ &     0.036$|{\bf p}_T|$
    \end{tabular}
    \\ 
    \end{tabular}
    \caption{{\bf Left:} Uncertainties of the Higgs boson \cite{Chen:2017bff} and {\bf Right:} pion
    momentum smearing parameters to be consistent with the Delphes configurations \texttt{delphes\_card\_CircularEE.tcl} \cite{Delphes:CEPC:link} for the CEPC/FCC-ee
    and \texttt{delphes\_card\_ILD.tcl} \cite{Abe:2010aa} for the ILC.}
\label{tab:smearing_parameters}
\end{table}

The pion momentum smearing is performed by randomly
sampling the azimuthal angle $\phi$ and the pseudo-rapidity
$\eta$ according to Gaussian distribution \cite{Rouge:2005iy,deFavereau:2013fsa}.
In addition, the
transverse momentum $|{\bf p}_T|$ is sampled with a Log-normal like distribution from Ref.~\cite{deFavereau:2013fsa},
\begin{eqnarray}
|{{\bf p}_T^{\rm rec}}|
=
  \exp
\left(
  \log |{{\bf p}_T}|
- \frac \epsilon 2\sqrt{1 + \frac{\sigma^2}{|{{\bf p}_T}|^2}}
\right),
\end{eqnarray}
with $\epsilon$ being a random number following a Gaussian
distribution centered in 0 with error 1 and $N$ a
normalization factor. For $\tau$ decay into $\rho^{\pm}$,
the reconstructed $\rho$ invariant mass is required to be
within the range of
$ 0.3~{\rm GeV} < \sqrt{p^2_\rho} < 1.2~$GeV.
The uncertainties of $(\phi, \eta, |{\bf p}_T|)$
for pions are summarized in the right part of
\gtab{tab:smearing_parameters}.

\begin{figure}[t!]
\centering
\includegraphics[scale = 0.28]{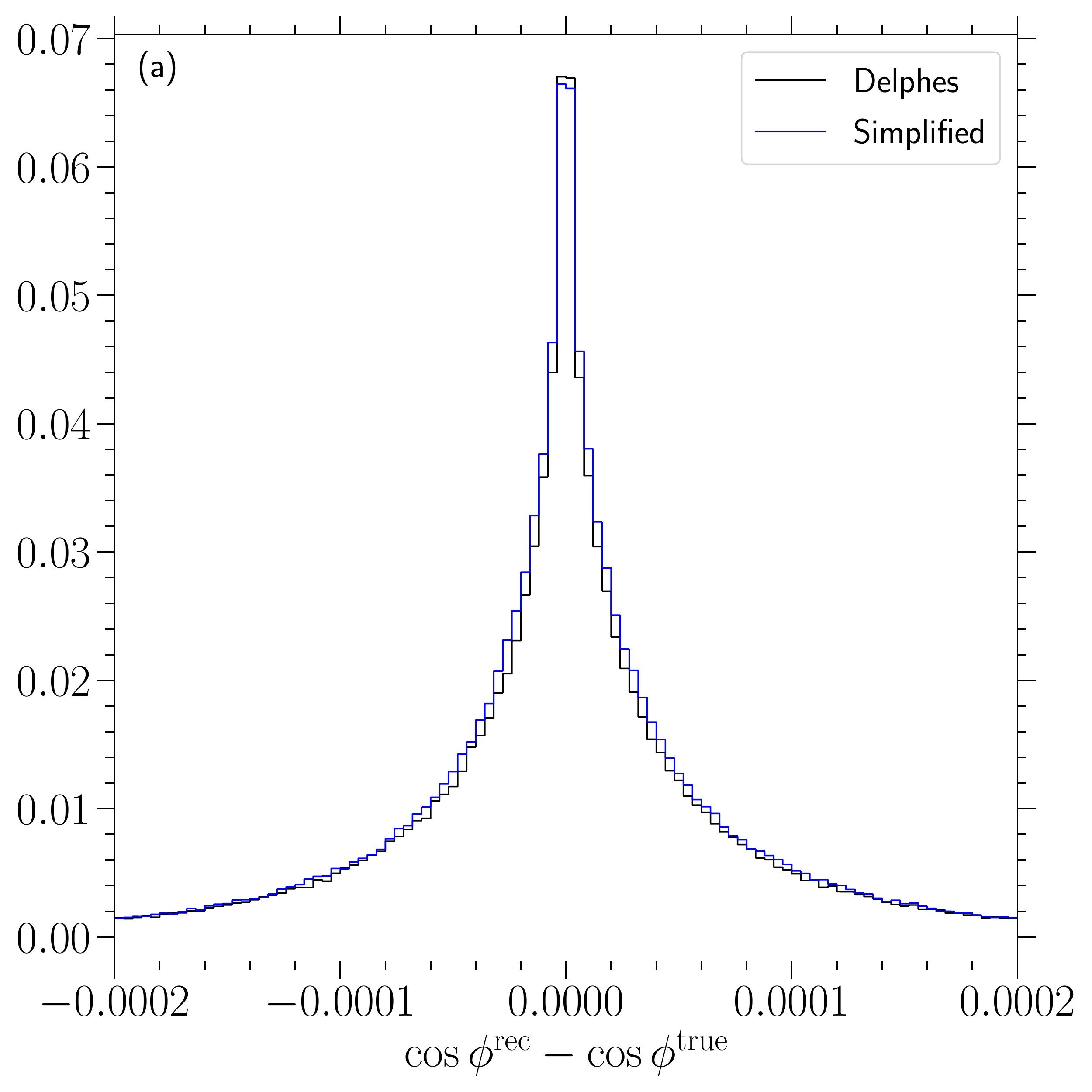}
\includegraphics[scale = 0.28]{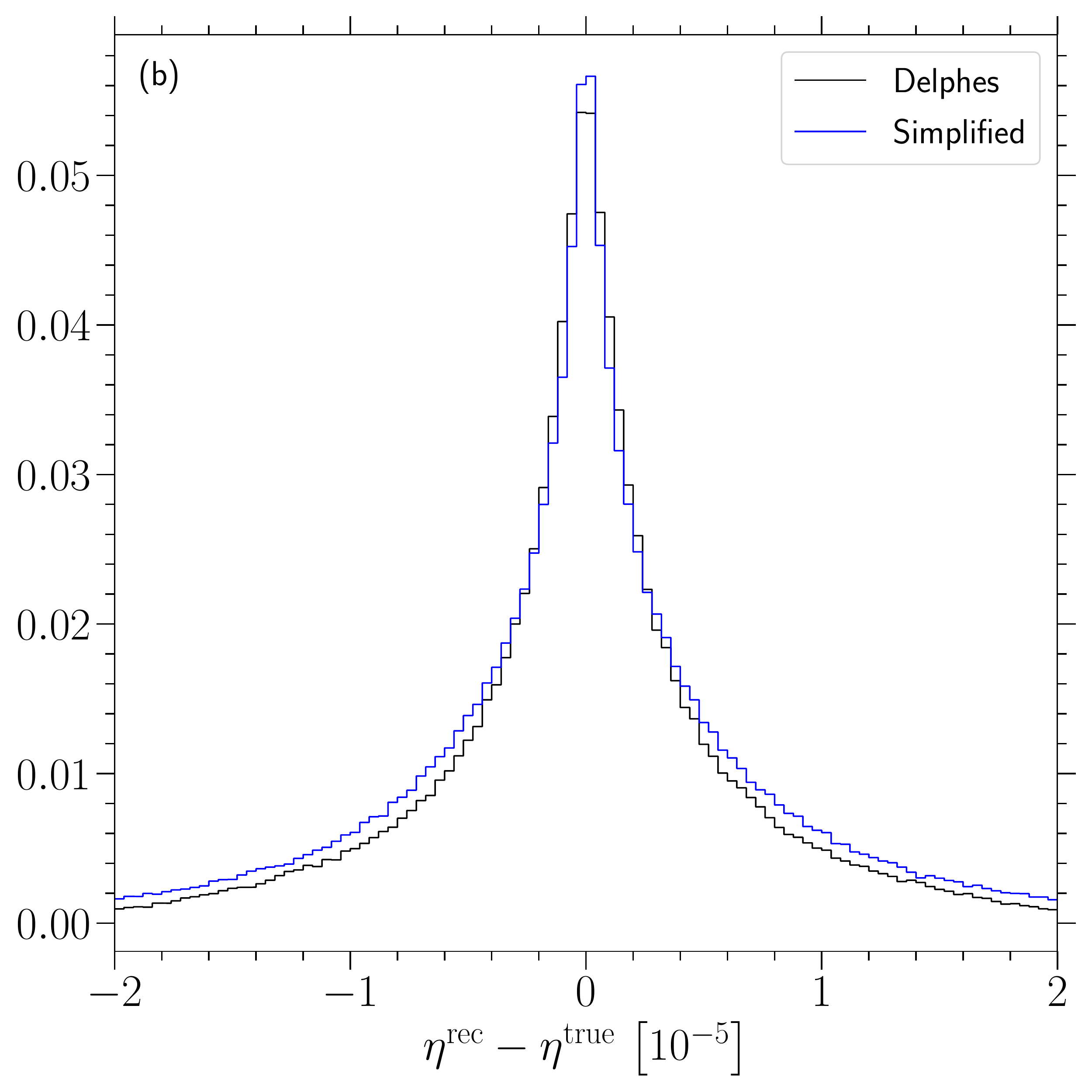}\\
\includegraphics[scale = 0.28]{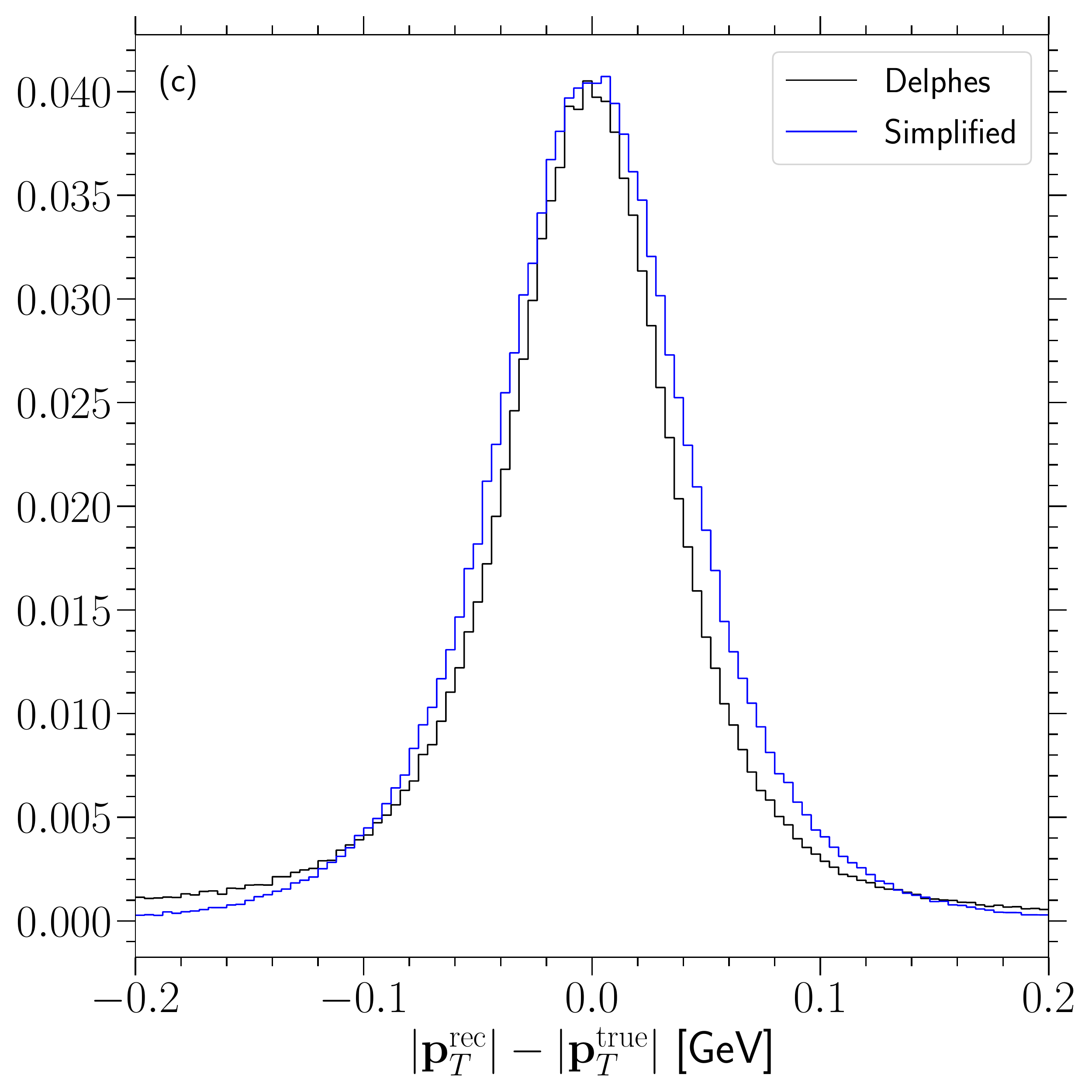}
\includegraphics[scale = 0.28]{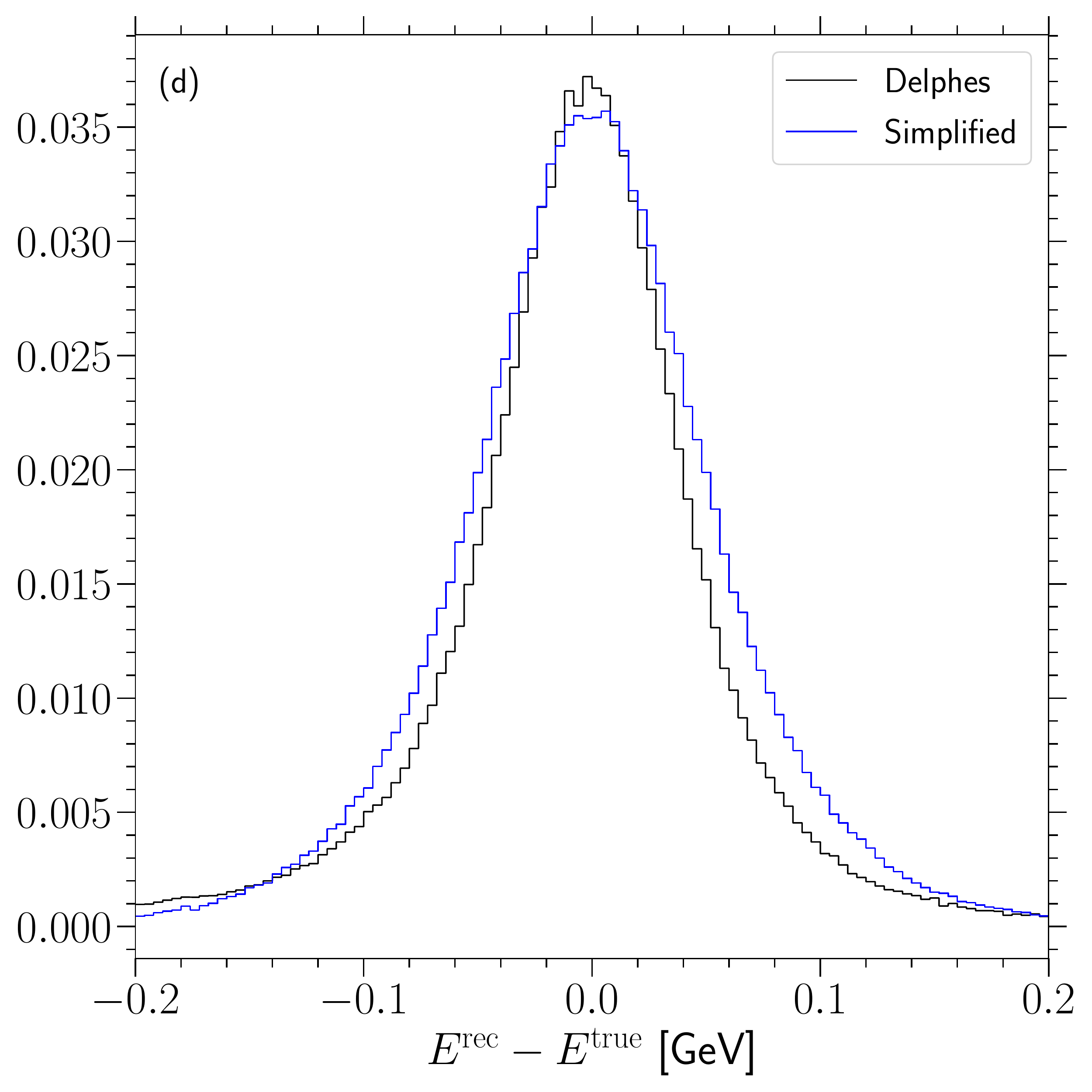}
\caption{The pion smearing effects simulated by
Delphes (black) and our simplified algorithm (blue). Notice that in the \gang{panel $(b)$} the numbers in the horizontal axis are multiplied by a factor $\times 10^5$ for better visualization.
}
\label{fig:delphes_vrs_ours}
\end{figure}

Although our simplified smearing algorithm is admittedly less sophisticated, our results are broadly compatible with those commonly adopted in the literature.
A complete analysis in momentum reconstruction and detailed cuts was performed in Refs.~\cite{deFavereau:2013fsa,Chen:2017bff}.
 For validation, we compare our smearing algorithm to
 the Delphes simulation with the configurations cards
\texttt{delphes\_card\_CircularEE.tcl}
\cite{Delphes:CEPC:link} for the CEPC/FCC-ee
\cite{Chen:2017yel} and
\texttt{delphes\_card\_ILD.tcl} \cite{Abe:2010aa}
for the ILC.
\gfig{fig:delphes_vrs_ours} shows the smeared
distributions of the pion kinematic variables simulated
with Delphes (black) vs our simplified smearing (blue).
We can see that the results of these two simulations
agree with each other quite well. In this work,
we take the simplified smearing algorithm for a fast
simulation.

To obtain the total number of expected events, one
needs to consider several branching ratios. 
First, the $Z$ boson can only be reconstructed if it
decays into either leptons or jets with $80\%$ of
branching ratio in total \cite{Zyla:2020zbs}. Also, since the decay branching ratio of Higgs
decaying into two $\tau$ leptons is $6.64\%$~\cite{Zyla:2020zbs}, only around
5.3\% of the actual Higgs events associated
with $Z$ production are available for the CP measurement.
Further suppression comes from the branching fraction
of the decay of $\tau$ into $\pi$ or $\rho$.
And we arrive at 7704 events at the CEPC, 7003 events at the
FCC-ee, and 4482 events at the ILC. Taking into account
the identification of $\tau$ jets and tagging of the
Higgs boson and other selection cuts \cite{Jeans:2018anq},
we obtain an overall efficiency,
$\epsilon = 0.145, 0.144, 0.142$ for $(\pi,\pi), (\pi,\rho)$
and $(\rho,\rho)$ decay modes,
respectively.

\begin{table}[H]
\centering
\begin{tabular}{c|c}
  Decay modes  & Branching ratio \\
\hline
  $Z \rightarrow $ vis. & 80\% \\
  $h \rightarrow \tau^+ \tau^-$ & 6.64\% \\
  $\tau \rightarrow \pi \nu_\tau$ & 10.82\% \\
  $\tau \rightarrow \rho \nu_\tau$ & 25.49\%
\end{tabular}
\qquad
\begin{tabular}{c|cc|cc|cc}
\multirow{2}{*}{\shortstack{$\tau$ decay \\ products}}  & \multicolumn{6}{c}{Number of Higgs decay events} \\
  &   \multicolumn{2}{c}{CEPC}   & \multicolumn{2}{c}{FCC-ee}  &  \multicolumn{2}{c}{ILC}   \\
  & before  & after &  before  & after & before  & after 
  \\ \hline
$(\pi,\pi)$     &   684     &    99     &   622     &   90     &   398      &   58   \\
$(\pi,\rho)$    &   3223    &   465     &   2930    &   423    &   1875     &   271   \\
$(\rho,\rho)$   &   3797    &   541     &   3451    &   491    &   2209     &   314
\end{tabular}
\caption{
{\bf Left: } 
    Branching fractions associated with the entire reaction. The values were obtained from~\cite{Zyla:2020zbs}.
{\bf Right:} 
    Expected event numbers at the CEPC~\cite{CEPCStudyGroup:2018ghi} with  the integrated luminosity $\mathcal{L} = 5.6~$ab$^{-1}$,
    FCC-ee~\cite{Abada:2019zxq} with $\mathcal{L}=5~$ab$^{-1}$ and ILC~\cite{Behnke:2013lya} with $\mathcal{L}=2~$ab$^{-1}$. The expected numbers of events before and after selection cuts are shown in the columns ``before'' and ``after'', respectively, with the overall cut efficiencies taken from Ref.~\cite{Jeans:2018anq}.}
\label{tab:event_expected}
\end{table}

The expected event numbers before and
after applying the selection efficiencies are shown
in \gtab{tab:event_expected} for comparison.
In total, roughly 1105, 1004, and 643 events of the
$h\to \tau^+\tau^-,\tau^\pm \to \pi^\pm/\rho^\pm \nu_{\tau^\pm}$ decay chains can be reconstructed
at the CEPC, FCC-ee, and ILC, respectively.

\subsection{Ambiguities in Momentum Reconstruction}

Experimentally, in order to reconstruct the
$\tau$ momentum, it is unavoidable to first obtain the
neutrino momentum which is not directly detectable.
With two neutrinos in the final state, we need to constrain
two 4-vector momenta. Since the Higgs momentum can be
fully reconstructed from the $Z$ boson counterpart, only one
neutrino momentum is independent due to energy-momentum
conservation. The 4 degrees of freedom can be constrained 
by the on-shell conditions of the two neutrinos and the
two $\tau$ leptons.

Unfortunately, the solutions have a two-fold ambiguity.
Since on-shell conditions are in quadratic forms, one
sign can not be uniquely fixed. For completeness,
we summarize the solution here in terms of the $\tau^-$
momentum defined in the Higgs rest frame,
\begin{eqnarray}
  {\bf p}_{\tau_-}
& = &
  \sqrt{p^2_h- 4 m^2_\tau}
  \left[
  \sin\theta_\tau 
  \left(
    \cos\phi_\tau
    {\hat {\bf n}}_1
  +
    \sin\phi_\tau
    {\hat {\bf n}}_2
  \right)
  \pm
  \cos \theta_\tau 
  {\hat {\bf n}}_3
  \right].
\label{eq:convenien_n_basis_prho}
\end{eqnarray}
The unit base vectors $\hat {\bf n}_i$ are constructed
in terms of the primary decay mesons,
$X_\pm \equiv \pi^\pm, \rho^\pm$,
\begin{eqnarray}
  {\hat {\bf n}}_1
=
  {\hat {\bf p}}_{X_+}
\,, \quad
  {\hat {\bf n}}_2
=  
    \frac
    {
      {\hat {\bf p}}_{X_-}
    -
     ({\hat {\bf p}}_{X_+}\cdot {\hat {\bf p}}_{X_-})
     {\hat {\bf p}}_{X_+}
    }
    {
      \sqrt
      {
        1
      -
        ({\hat {\bf p}}_{X_+}\cdot {\hat {\bf p}}_{X_-})^2
      }
    },
\qquad
  {\hat {\bf n}}_3
=
  \frac
  {{\hat {\bf p}}_{X_+} \times {\hat {\bf p}}_{X_-}}
  {
    \sqrt
      {
        1
      -
        ({\hat {\bf p}}_{X_+}\cdot {\hat {\bf p}}_{X_-})^2
      }
  }\,.
\end{eqnarray}
The first base vector $\hat{\bf n}_1$ is along the
momentum of $\pi^+$ or $\rho^+$ while the third one
$\hat{\bf n}_3$ is perpendicular to the momentum of
both primary mesons. Finally, $\hat{\bf n}_2$ is simply
the one perpendicular to both $\hat{\bf n}_1$ and
$\hat{\bf n}_3$. The polar angles
of the $\tau$ momentum can be reconstructed as,
\begin{subequations}\label{eq:analytic_pt}
\begin{eqnarray}
  \sin\theta _\tau
  \cos \phi_\tau
 & = & 
  \frac{m^2_\tau + m^2_X - m_h E_{X_{+}}}{|{\bf p}_{X_{+}}|\sqrt{m^2_h - 4 m^2_\tau}},
\\
  \sin\theta_\tau
  \sin \phi_\tau
& = &
    \frac{m_h E_{X_{-}} - m^2_\tau - m^2_{X_-}}{|{\bf p}_{X_{-}}||s_{X_- X_+}|\sqrt{m^2_h - 4 m^2_\tau}}
  +
    \frac{m_h E_{X_{+}} - m^2_\tau - m^2_{X_+} }{|{\bf p}_{X_{+}}||s_{X_- X_+}|\sqrt{m^2_h - 4 m^2_\tau}}
    c_{X_- X_+}\;,
\qquad
\end{eqnarray}
\end{subequations}
where $(s_{X_- X_+}, c_{X_- X_+}) \equiv (\sin \theta_{X_- X_+}, \cos \theta_{X_- X_+})$
and $\theta_{X_- X_+}$ is the angle between the
momentum of $X_+$ and $X_-$. 

However, in \geqn{eq:convenien_n_basis_prho} the
$\pm$ sign in front of $\hat {\bf n}_3$ reflects the fact
that both solutions obey all the constraints from
energy-momentum conservation and the correct solution
cannot be unambiguously obtained. 
This sign ambiguity can significantly decrease the
CP sensitivity, especially for the neutrino azimuthal
angle distribution.
Using momentum conservation, the result in \geqn{eq:azimuth_diff_definition} for $\tan\delta \phi_\nu$ can be written in the same form by substituting ${\bf p}_\nu$ by ${\bf p}_{X^\pm}$, hence, $\tan \delta \phi_\nu \propto {\hat{\bf p}}_{\tau^-} \cdot ({\bf p}_{X^+}\times {\bf p}_{X^-}) = \pm \cos \theta_\tau/\sqrt{1-({\hat {\bf p}}_{X_+}\cdot {\hat {\bf p}}_{X_-})^2 }$.
In other words, $\delta \phi_\nu$ can have both positive
and negative solutions with the same magnitude.
This would not be a big problem for the symmetric
distribution of $\delta \phi_\nu$ around its origin,
such as those curves in \gfig{fig:comparison_vars}
with $\Delta = 0^\circ$. But it causes significant issues
for other $\Delta$ values and effectively flattens the curve for $\Delta = \pm 45^\circ$.

This ambiguity can be solved by measuring other decay
information. An especially useful quantity 
is the impact parameter \cite{Kuhn:1993ra, Jeans:2015vaa}, the minimum
distance of charged meson trajectory  to the
$\tau$ leptons production point. The impact
parameter measurement essentially removes the two-fold
ambiguity for the $\tau$ Yukawa CP measurement
at future lepton colliders \cite{Desch:2003mw,Rouge:2005iy}.
A more recent study with spatial resolution of 5 $\mu$m
can be found in Refs.~\cite{Chen:2017bff,Jeans:2018anq}.

Another ambiguity comes from the detector resolutions.
The $\tau$ momentum is reconstructed from the smeared
Higgs and meson momentum. This reconstruction is
realized with energy and momentum conservation,
assuming narrow width approximation for the $\tau$ momentum,
$p_{\tau^\pm}^2 = m_\tau^2$. Both smearing and finite
width could lead to nonphysical solutions in \geqn{eq:analytic_pt}, for example $\sin \theta_\tau \sin \phi_\tau> 1$. For those events, we follow a similar procedure introduced in Ref.~\cite{Chen:2017bff}. 
We try to find the solution for the (anti-)neutrino momenta optimally consistent
with all the information we have on each event (including four-momentum conservation) by minimizing the 
function
\begin{eqnarray}
  \chi^2_{\rm rec}
=
  \sum_{i=0}^3 
  \left(
    \frac{(p^{\rm rec}_{\tau^+})_i + (p^{\rm rec}_{\tau^-})_i - (p^{\rm rec}_h)_i}
    {\sigma_h}
    \right)^2
+
  \left(
  \frac{(p^{\rm rec}_{\tau^+})^2 - m_\tau^2}{\sigma_\tau}
  \right)^2
+
  \left(
  \frac{(p^{\rm rec}_{\tau^-})^2 - m_\tau^2}{\sigma_\tau}
  \right)^2,
\end{eqnarray}
where $i = 0, \cdots, 3$ runs over the 4-momentum components
of each particle momentum.
We adopt the uncertainties as $\sigma_h = 4.0$ GeV and $\sigma_\tau = 0.2$ GeV \cite{Chen:2017bff}.
The $\chi^2_{\rm rec}$ function is minimized over
the 6 kinematic parameters of the unmeasured neutrino
momentum:
the pseudo-rapidity, azimuthal angle, and absolute value of
the momentum for both neutrino/anti-neutrino. Then the
$\tau$ momentum is then obtained with energy momentum
conservation, $p^{\rm rec}_{\tau^\pm} = p_{\nu^\pm} + p^{\rm rec}_{X^\pm}$. The best fit at the minimum of $\chi^2_{\rm rec}$ approximates the physical solution.
We keep the event if the minimum solution is consistent
with the mass cuts. Otherwise, the event is discarded. 
\begin{figure}[H]
\centering
\includegraphics[scale = 0.18]{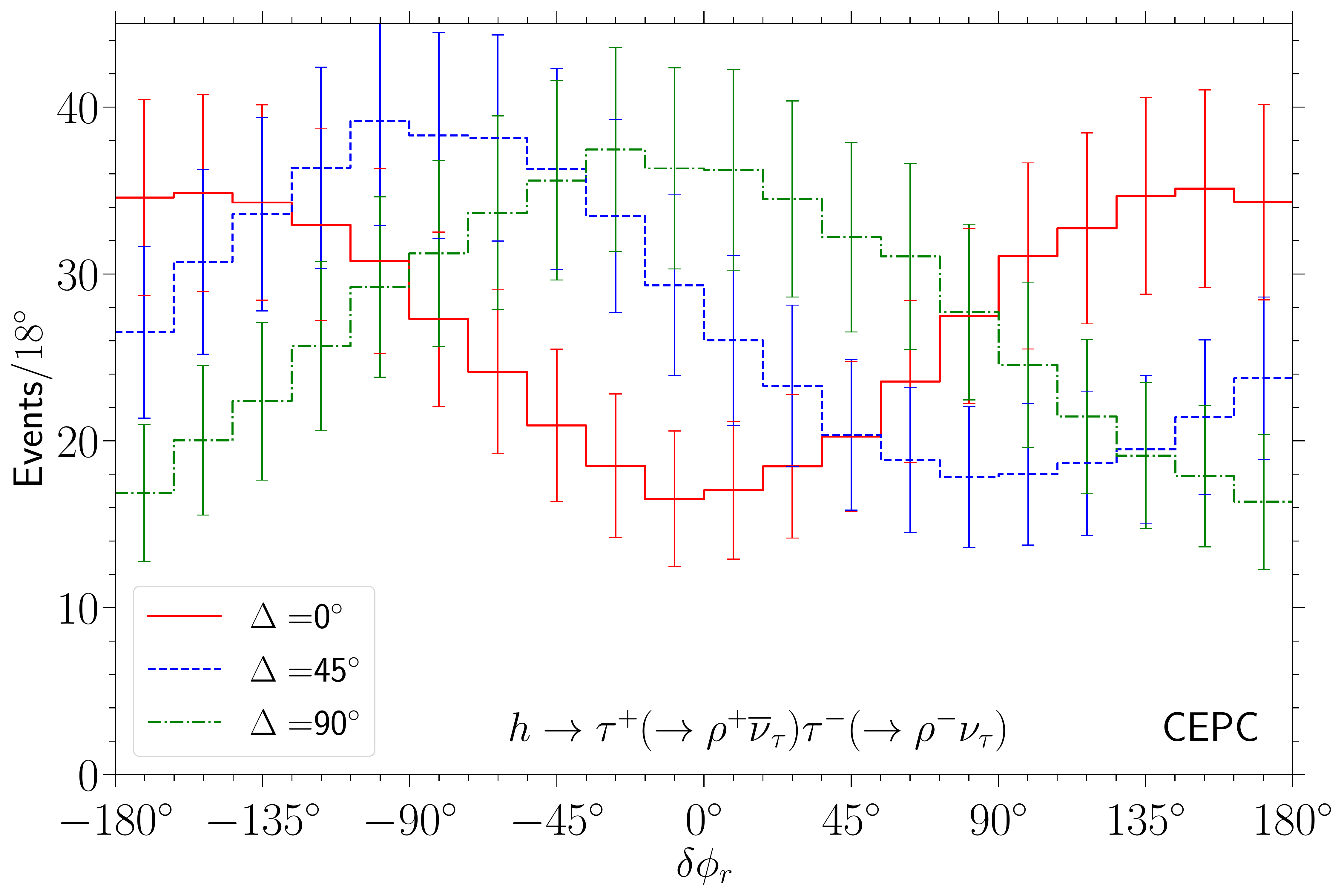}
\includegraphics[scale = 0.18]{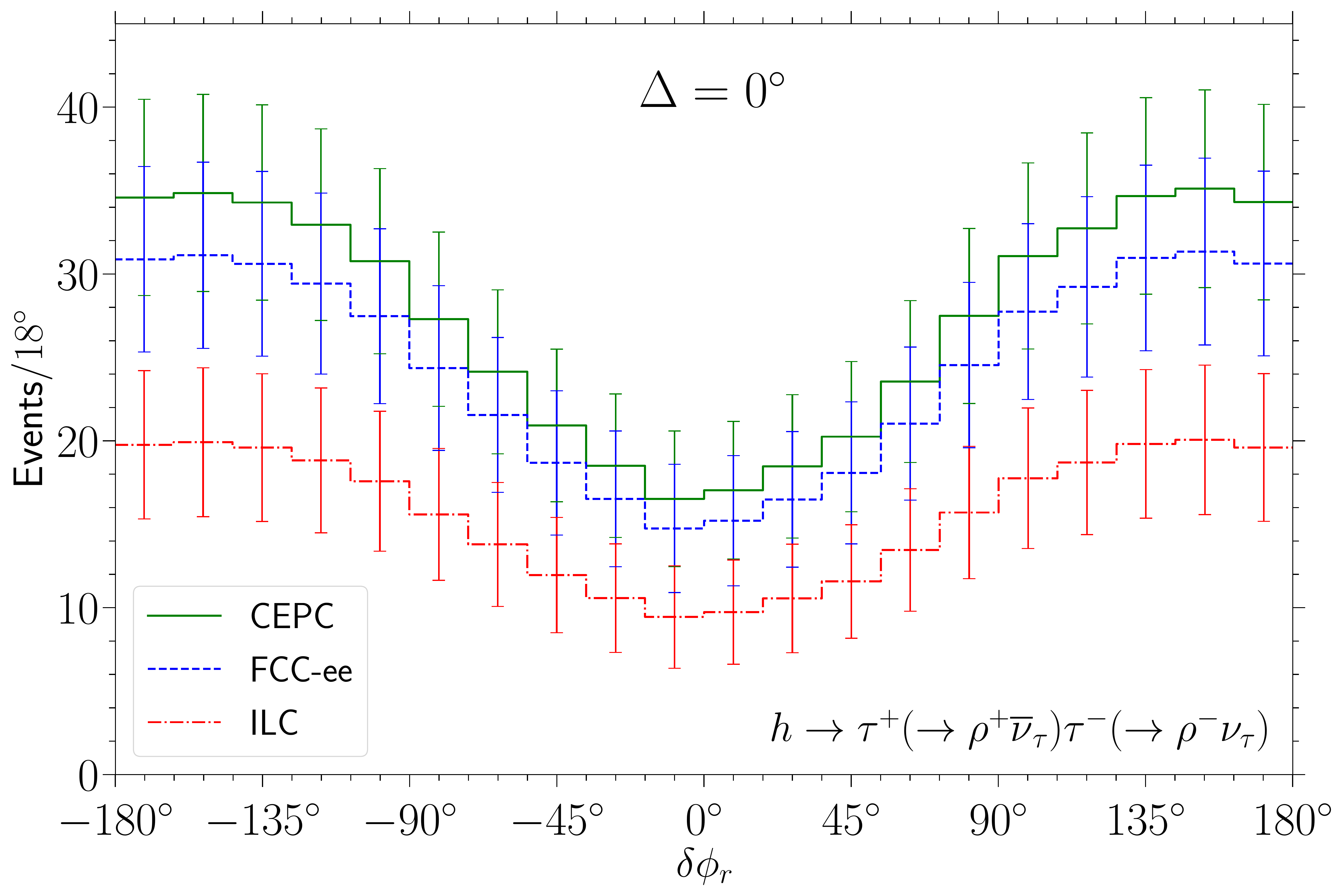}
\caption{Smeared differential distributions of $\delta\phi_r$ in the decay mode $ h\rightarrow \tau^+(\rightarrow\rho^+ \overline \nu_\tau)\tau^-(\rightarrow\rho^- \nu_\tau)$.
      {\bf Left:}
      The distribution for CP phases $\Delta = 0^\circ$ (red), $\Delta = 45^\circ$ (blue), $\Delta = 90^\circ$ (green) at the CEPC;
     {\bf Right:}
      A comparison of the distributions at the
      CEPC (red), FCC-ee (blue), and ILC (green) for $\Delta=0^\circ$. In both panels, the error bars indicate the statistical uncertainties.}
\label{fig:event_dist_example}
\end{figure}

The final result of the differential distribution for
the $ h\rightarrow \tau^+(\rightarrow\rho^+ \overline \nu_\tau)\tau^-(\rightarrow\rho^- \nu_\tau)$ process is  plotted in
\gfig{fig:event_dist_example}.
The left panel shows
the differential distributions for $\Delta = 0^\circ$ (red),
$\Delta = 45^\circ$ (blue), and $\Delta = 90^\circ$
(green), respectively.
Being divided into 20 bins \cite{Harnik:2013aja,Jeans:2018anq},
there are $25 \sim 35$ events in each bin on average.
The corresponding statistical uncertainty at the level of
$17\% \sim 20\%$ is much smaller than the oscillation amplitude,
$\pi^2 / 16 \approx 62\%$. The event rate at the CEPC are
large enough to constrain
the modulation pattern as elaborated in \gsec{sec:CPV}.
The right panel shows the spectrum at the three future
candidate lepton colliders, CEPC (red), FCC-ee (blue),
and ILC (green), respectively, for comparison.
While CEPC and FCC-ee have comparable spectrum, ILC
has much lower event rate and hence larger fluctuations.

It is interesting to see that for $\Delta = 90^\circ$,
the differential distribution of $\delta \phi_r$
has only $\cos\delta\phi_r$ but no $\sin\delta\phi_r$
in \geqn{eq:angular_distribution_polarimeter}.
In other words, the observable that we
measure has only CP conserving contribution that
does not change under CP
transformation. However, the distributions in the
left panel of \gfig{fig:event_dist_example} show that
the difference between $\Delta = 0^\circ$ and
$\Delta = 90^\circ$ is maximal. This is because
$\cos 2 \Delta = \pm 1$ take the two extreme
values with opposite signs.

\subsection{Discovery Potential and Sensitivity of the CP Phase}
\label{sec:CPV}

To evaluate the CP measurement sensitivities,
we adopt a $\chi^2$ function defined according
to the Poisson distribution,
\begin{eqnarray}\label{eq:chi2_def}
\chi^2
\equiv
  \sum_i
    2(N_i^{\rm test} - N_i^{\rm true})
  +
   2 N_i^{\rm true}\log(N_i^{\rm true}/N_i^{\rm test}),
\end{eqnarray}
where $i = 1, \cdots, 20$ runs over all the 20 bins of
the $\delta \phi_r$ differential distribution. Since
we are studying the projected sensitivity at future
lepton colliders, there is no real data available yet.
Instead, we simulate the measurement with some assumed
true values of the CP phase $\Delta$ to produce a set of
pseudo-data $N^{\rm true}_i$ and then fit these pseudo-data
with some test values $N^{\rm test}_i$. The event
numbers $N^{\rm true}_i$ and $N^{\rm test}_i$ are
functions of the true value $\Delta^{\rm true}$ and
$\Delta^{\rm test}$, respectively.

\begin{figure}[t]
\centering
\includegraphics[scale = 0.18]{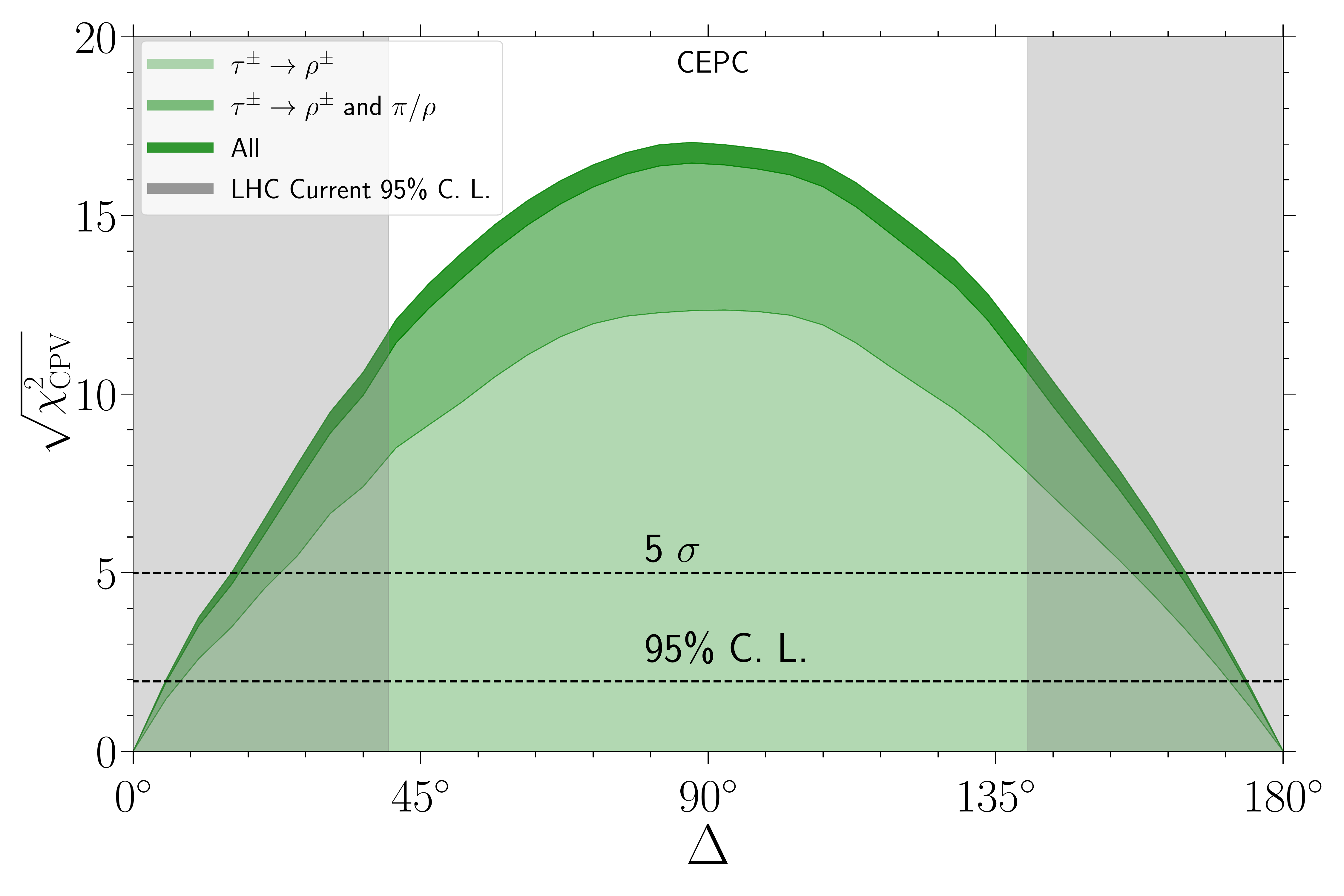}
\includegraphics[scale = 0.18]{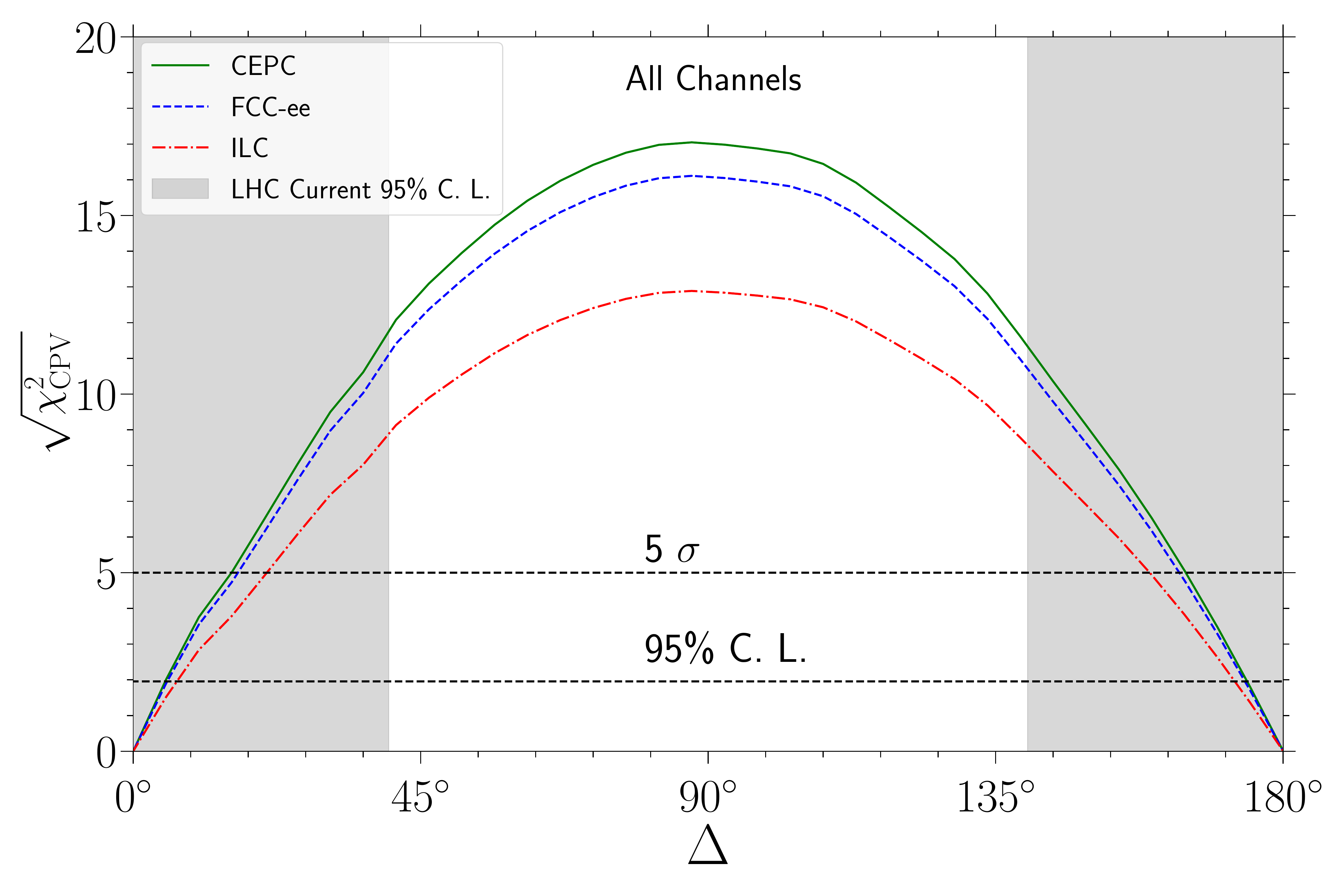}
\caption
{
{\bf Left: }
The CP phase discovery potential at the CEPC
for $\kappa_\tau=1$.
The green shaded regions represent the results from
various decay modes: only $(\rho,\rho)$ (light green),
$(\rho,\rho) + (\pi,\rho)$ (green) and
the fully combined one
$(\rho,\rho) + (\pi,\rho) + (\pi, \pi)$ (dark green),
with the boundaries
describing the $\sqrt{\chi^2_{\rm CPV}}$ values
according to Eq.~\eqref{eq:chi-square-CPV} given
$\Delta^{\rm{true}}$. The black dashed lines with
$\sqrt{\chi_{\rm{CPV}}^2}= 1.96,5$ mark the
sensitivities at $95\%$ C.L. and $5\sigma$, respectively.
{\bf Right: }
Sensitivity of all the channels at the
CEPC (green), FCC-ee (blue), and ILC (red). In both panels,
the region outside of the gray bands are excluded at
95\% C.L. by the current CP measurement at the LHC
\cite{CMS:2020rpr}.
}
    \label{fig:CPVsensitivity_plot}
\end{figure}

The discovery ability of a nonzero CP phase can be
parametrized as the smaller one of the two $\chi^2$
values between the given $\Delta^{\rm true}$ and the CP
conserving cases $\Delta^{\rm test} = 0^\circ$ or
$\Delta^{\rm test} = 180^\circ$,
\begin{eqnarray}
\label{eq:chi-square-CPV}
\chi^2_{\rm CPV}(\Delta^{\rm true})
\equiv
  {\rm min}[\chi^2(\Delta^{\rm true}, \Delta^{\rm test} = 0^\circ),\chi^2(\Delta^{\rm true}, \Delta^{\rm test} = 180^\circ)].
\end{eqnarray}
\gfig{fig:CPVsensitivity_plot} shows the
$\sqrt{\chi^2_{\rm CPV}}$ distribution as a function of
$\Delta$  assuming  $\kappa_\tau=1$. 
The sensitivities for the di-$\tau$ decay into $(\rho,\rho)$,
$(\rho,\rho) + (\pi,\rho)$, and the full combination
$(\rho,\rho) + (\pi,\rho) + (\pi, \pi)$ are depicted 
in light green, green, and dark green regions, respectively.
For different decay channels, the differential distributions
have the same amplitude $\pi^2/16$ as indicated in 
\geqn{eq:angular_distribution_polarimeter}. So
the main difference in the sensitivities is due to the
event rates: the branching ratio of the
$\tau \rightarrow \pi \nu_\tau$ is only $10.8\%$,
in comparison with the $25\%$ for the
$\tau \rightarrow \rho \nu_\tau$ channel.
As indicated by the black dashed lines, 95\% of
the values of $\Delta$ can be tested above $95\%$ C.L.
and 82\% of the parameter space can be tested at
even more than $5\sigma$.
The sensitivity peaks at $\Delta = \pm 90^\circ$
where \geqn{eq:angular_distribution_polarimeter}
takes the most different value from that of
$\Delta = 0^\circ$ or $180^\circ$ with more than
$10\,\sigma$ significance. In the right panel, we also 
show the comparison of the sensitivities at the CEPC
(green), FCC-ee (blue),
and ILC (red). As expected, the CEPC has the highest
sensitivity due to the higher number of events.

\begin{table}[ht]
\centering
\begin{tabular}{c|ccc}
& 68\% C.L. for $m=1$ & 95\% C.L. for $m=1$ & 95\% C.L. for $m=2$ \\
\hline
CEPC   & 2.9$^\circ$ & 5.6$^\circ$ & 7.0$^\circ$ \\
FCC-ee & 3.2$^\circ$ & 6.3$^\circ$ & 7.8$^\circ$ \\
ILC    & 3.8$^\circ$ & 7.4$^\circ$ & 9.3$^\circ$
\end{tabular}
\caption{The CP phase precision at the CEPC, FCC-ee, and ILC for $m$ parameter(s)}.
\label{tab:confidence_intervals}
\end{table}

For completeness, \gtab{tab:confidence_intervals}
summarizes the expected precision of the $\Delta$ measurement
at  future lepton colliders at 68\% C.L. and 95\% C.L.
for $m=1$ parameter ($\Delta$), or
$m = 2$ parameters ($\Delta$ and $\kappa_\tau$).
Notice that our estimation at $1\,\sigma$ level is
slightly better than the $4.4^\circ$ with $1\,\mbox{ab}^{-1}$ in only the $\tau \rightarrow \rho \nu_\tau$ decay channel
\cite{Harnik:2013aja} or $4.3^\circ$ with
$2\,\mbox{ab}^{-1}$ in
both $\tau$ decay channels \cite{Jeans:2018anq} at the ILC. For the CEPC, our
result is the same as the $2.9^\circ$ in Ref.~\cite{Chen:2017bff}. Notice that in addition to
the two mesonic decay channels, the leptonic decay channel
$\tau \rightarrow \ell \nu \bar \nu$ is also considered
in Ref.~\cite{Chen:2017bff} with
the matrix element based observable that is
different from our polarimeter $\delta \phi_r$.
We can clearly see from \gtab{tab:confidence_intervals} that
the future lepton colliders can differentiate the CPV
scenario from the CP-conserving one very well.

\section{Prospects of Constraining New Physics}\label{sec:new_phys_BAU}

As the aforementioned analysis shows, there remains significant potential for discovering CP violation in the $h\to\tau^+\tau^-$ decay at prospective future lepton colliders. We now draw the connection with the lepton flavored EWBG scenario, following the treatment given in Ref.~\cite{Guo:2016ixx} for concrete illustration (see Refs.\cite{Chiang:2016vgf,deVries:2018tgs,Fuchs:2020uoc,Xie:2020wzn}).This discussion exemplifies future lepton colliders are not
only precision machines but can also make an $\mathcal O(1)$
measurement of BSM physics effects.

\subsection{Two Higgs Doublet Model}

The set up in Ref.~\cite{Guo:2016ixx} relies on the type III Two Higgs Doublet Model (THDM)~\cite{Barger:1989fj,Branco:2011iw}, wherein the
two scalar doublet fields before EWSB are denoted as $\Phi_{1,2}$.
Both neutral scalars inside $\Phi_{1,2}$ acquire
nonzero VEVs, $v_1$
and $v_2$, respectively, with
$v \equiv \sqrt{v_1^2 + v_2^2} = 246$ GeV.
The neutral components can mix with each other to
form three neutral massive scalar fields after
one neutral Goldstone boson is eaten by the $Z$ boson. We assume a CP-invariant scalar potential,
namely, only the real parts of the two neutral
scalars can mix with each other but not with
the imaginary parts,
\begin{eqnarray}
  H
\equiv 
  c_\alpha {\rm Re}[\Phi^0_1] + s_\alpha {\rm Re}[\Phi^0_2]
,
\quad
  h
\equiv
- s_\alpha {\rm Re}[\Phi^0_1] + c_\alpha {\rm Re}[\Phi^0_2],
\quad
  A
\equiv
  -s_\beta {\rm Im}[\phi_1^0] + c_\beta{\rm Im}[\phi_2^0],
\end{eqnarray}
where $s_\alpha \equiv \sin \alpha$,
$c_\alpha \equiv \cos \alpha$, 
$\tan \beta \equiv v_2/v_1$, and Re and Im denote the real and imaginary parts, respectively. Note that $\alpha$
is the mixing angle from the neutral scalar mass matrix
diagonalization. The neutral particle masses are ordered
as $m_H, m_A > m_h \approx 125$ GeV, so that $h$ is the SM-like Higgs
boson.

In the Type-III THDM, the Yukawa interaction for each doublet field has the same structure as the SM Yukawa interaction;
\begin{eqnarray}
  \mathcal{L}_Y
=
-
  \overline L Y_1 \ell_R \Phi_1
-
  \overline L Y_2 \ell_R \Phi_2
  +{\rm h.c.}
\label{eq:yukawa_before_EL_weak}
\end{eqnarray}
In this way, both Higgs doublets can contribute
its neutral components to couple with the $\tau$
lepton \cite{Berge:2015nua},
\begin{eqnarray}
  -\frac{m_\tau}{v}\overline{\tau}_L\tau_R
  \left[
    \left(s_{\beta -\alpha}+\frac{N_{\tau\tau}}{m_\tau}c_{\beta -\alpha}\right)h
  +
    \left(c_{\beta -\alpha}-\frac{N_{\tau\tau}}{m_\tau}s_{\beta -\alpha}\right)H
  +
  iA N_{\tau\tau}
  \right],
\end{eqnarray}
where $N_{\tau\tau}$ is a complex parameter
related to the matrix elements of
$Y_{1,2}$. Following the parametrization of
\geqn{eq:interaction}, the $\tau$ Yukawa coupling
becomes 
\begin{eqnarray}
  \kappa_\tau (\cos \Delta + i \sin \Delta)
=
  s_{\beta -\alpha}
+ \frac{N_{\tau\tau}}{m_\tau} c_{\beta -\alpha}.
\end{eqnarray}
Notice that CP violation arises due to the
imaginary part of $N_{\tau \tau}$. 
Moreover, for the
particular texture, $Y_{j,22} = Y_{j,23} = 0$,
$Y_{1,33} = Y_{2,33} = Y_{33}$ and
$Y_{1,32} = r_{32}Y_{2,32}$, one can write the
imaginary part of the Jarskog invariant $J_A$ of the Yukawa
interaction in \geqn{eq:yukawa_before_EL_weak} as,
\begin{eqnarray}
\label{eq:Jarlskog}
  \mathrm{Im} [J_A]
=
- \mathrm{Im} [r_{32}]|Y_{2,32}|^2
=
  \frac{2 m^2_\tau}{v^2c_{\beta -\alpha}}
  \kappa_\tau \sin \Delta \,.
\end{eqnarray}
It is the imaginary part of the Jarlskog invariant that
controls the size of the BAU in early universe through
lepton flavored baryogenesis \cite{Guo:2016ixx}.
Rewriting \geqn{eq:Jarlskog} gives
\begin{equation}
\sin\Delta
= \frac{v^2 c_{\beta -\alpha}}{2 m^2_\tau \kappa_\tau} \mathrm{Im} [J_A].
\end{equation}
Thus, one may connect the $\tau$ Yukawa CP
phase $\Delta$, which can be measured at future
lepton colliders, with CPV source for baryogenesis
during the era of EWSB in the early universe.

\subsection{Sensitivity to the Baryogenesis Scenario}

\begin{figure}[H] 
    \centering
    \includegraphics[scale = 0.3]{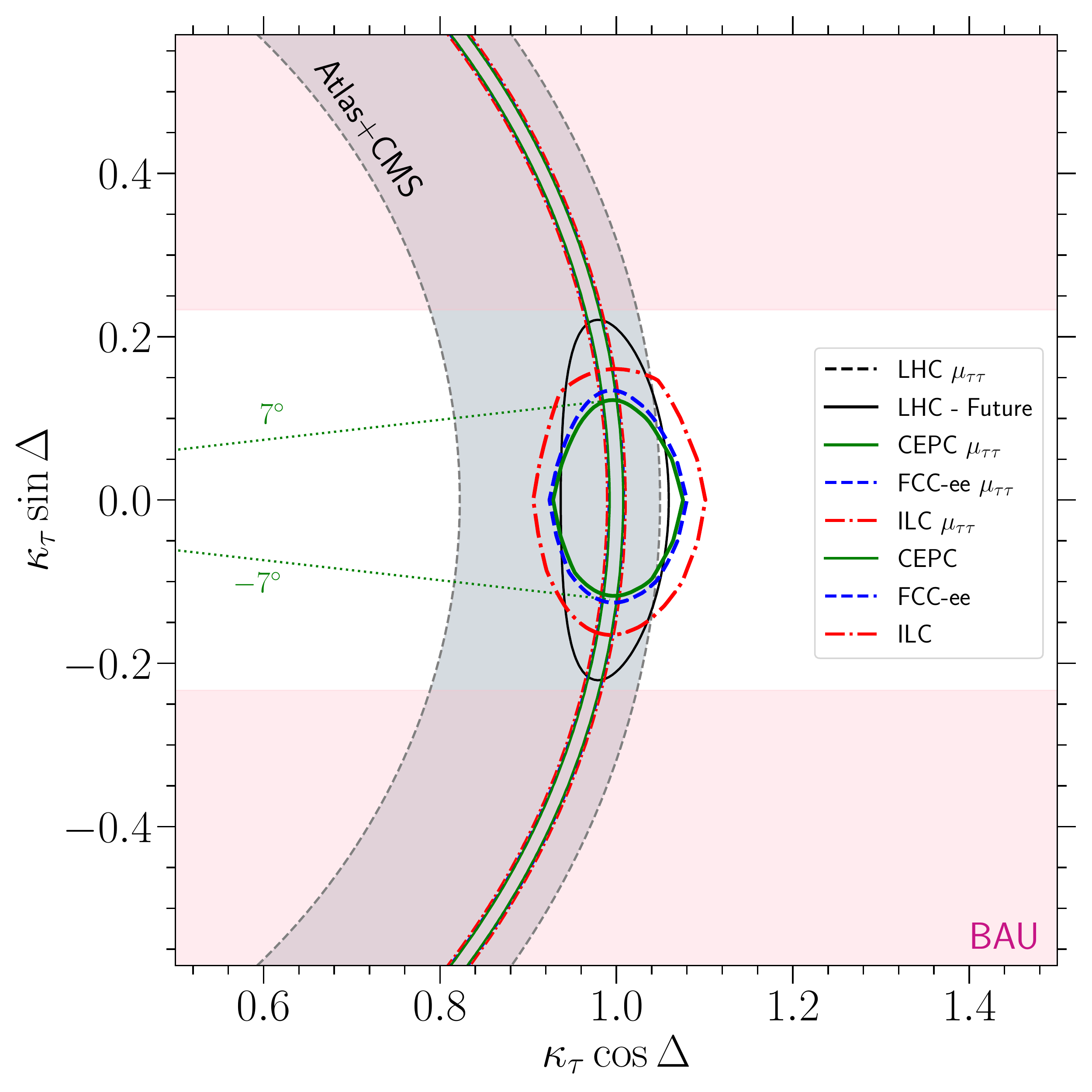}
    \includegraphics[scale = 0.3]{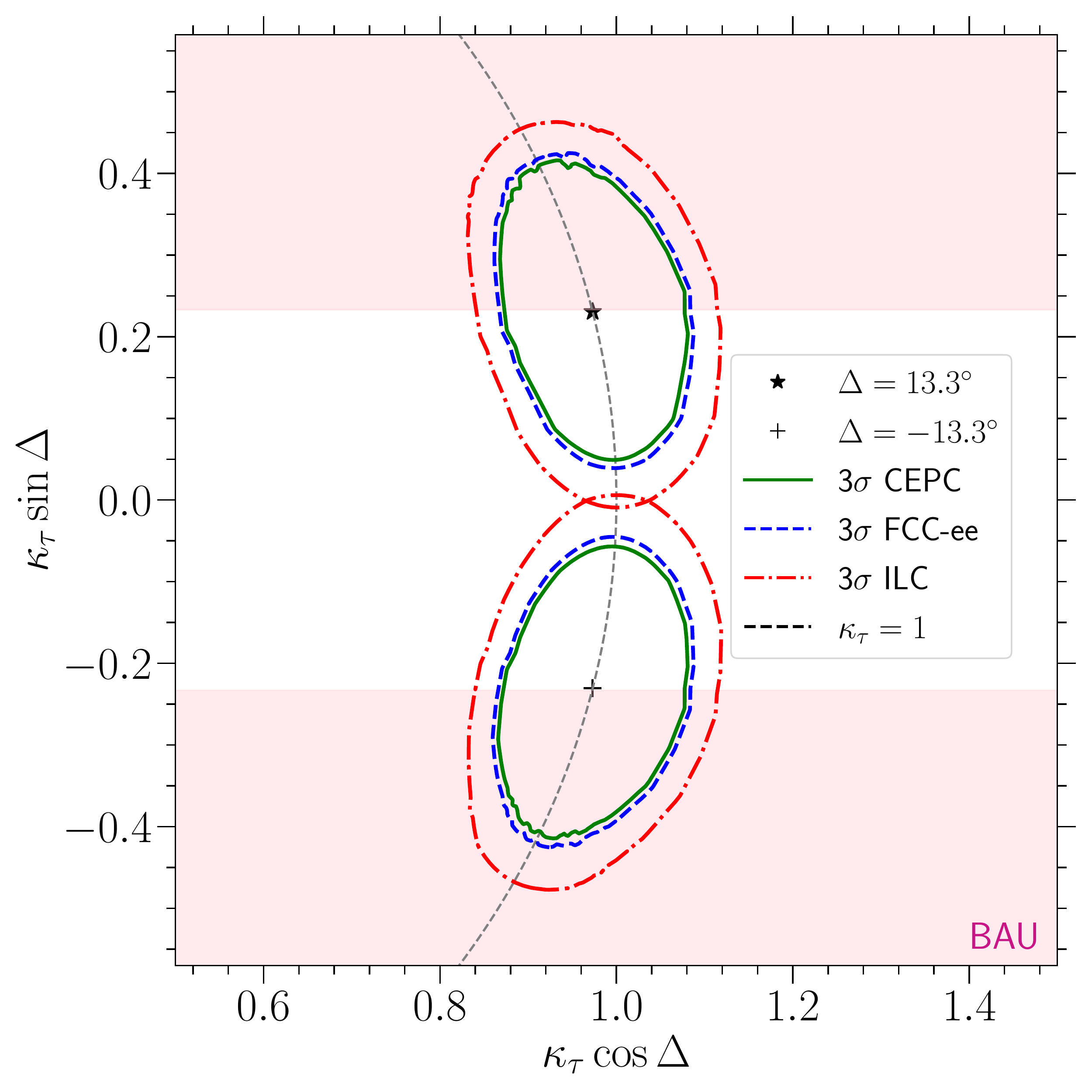}
    \caption
    {
    {\bf Left:}
      The 95\% C.L. constraints on the $\tau$ Yukawa coupling 
      at the CEPC (green), FCC-ee (blue), and ILC (red) assuming the true value
      $\Delta^{\rm true} = 0^\circ$ and $\kappa^{\rm true}_\tau = 1$. 
      The gray band gives 95\% C.L. constraints from the current LHC signal strength measurements \cite{Aaboud:2018pen,CMS:2020dvp} while
      the black contour denotes the expected 95\% C.L. constraint from the combined measurements of $\mu_{\tau\tau}$~\cite{CMS:2013xfa,TheATLAScollaboration:2014ewu} and $\Delta$~\cite{Chen:2017nxp} at the HL-LHC.
    {\bf Right: }
      The 3$\sigma$ contours for each collider assuming central values $\Delta^{\rm true} = \pm13.3^\circ$ and $\kappa^{\rm true}_\tau = 1$ corresponding to the minimum $|\kappa_\tau \sin\Delta|$ compatible with the BAU.
    }
    \label{fig:sensitivity_plot}
\end{figure}

To make this connection concrete, we plot in Fig.~\ref{fig:sensitivity_plot} the 95\% C.L. constraints on 
$\kappa_\tau\cos\Delta$ and $\kappa_\tau\sin\Delta$
from present and future collider probes and from lepton flavored EWBG.
For generality, we also set $\kappa_\tau^{\rm test}$ to be free to obtain a full picture on a two-dimensional plot.
The CP sensitivity is then depicted as the
contours around the true value
$\Delta^{\rm true} = 0^\circ$ and $\kappa^{\rm true}_\tau = 1$
in the left pannel, with the green, blue (dashed),
and red (dot dashed) contours
indicating the 95\% C.L. sensitivities.
The green dotted lines
from the origin $\kappa_\tau = 0$ are added to show that
the contour size corresponds to roughly $7^\circ$ at
95\% C.L. Consistent with the previous
observation, the CEPC and FCC-ee have comparable
precision while that of the ILC is slightly weaker due to different
luminosities. For all three cases, the pink region
allowing for successful explanation of BAU is outside
the 95\% C.L. contour. In other words, the lepton flavored BAU mechanism as given in Ref.~\cite{Guo:2016ixx}
could be excluded at better than than 95\% C.L. .
For comparison, we also show the projected $\tau$ Yukawa CP
measurement at the High Luminosity (HL-)LHC \cite{Chen:2017nxp}
with the integrated luminosity of $3\,\mbox{ab}^{-1}$ 
as the black contour, which will be further elaborated
below. It is clear that even with
the HL-LHC, the THDM BAU mechanism can only be tested with
barely 95\% C.L. The CP measurements
at future lepton colliders can significantly improve
the situation.

We also include the constraints from the
measurement of the $h\to\tau\tau$ decay signal strength
$\mu_{\tau\tau}$, which is proportional to $\kappa_\tau^2$.
The current data at the LHC indicate
$\mu_{\tau\tau}=1.09^{+0.35}_{-0.30}$ at ATLAS
\cite{Aaboud:2018pen} and 
$\mu_{\tau\tau}=0.85^{+0.12}_{-0.11}$ at CMS 
\cite{CMS:2020dvp}, which are depicted as the gray region.
In other words, the current measurement at LHC is still
quite crude with at least 10\% uncertainty.
At the HL-LHC, the $1\,\sigma$ uncertainty of
$\mu_{\tau\tau}$ can be further improved to $5\%$ 
\cite{CMS:2013xfa,TheATLAScollaboration:2014ewu},
which is further combined with the CP
measurement \cite{Chen:2017nxp} that is shown as  the
black contour.
The future lepton colliders can significantly improve the
sensitivities to 0.8\% at the CEPC~\cite{Yu:2020bxh},
0.9\% at the FCC-ee~\cite{Abada:2019zxq},
and 1.09\% at the ILC~\cite{Yu:2020bxh},
which are shown as the rings in the left panel
of \gfig{fig:sensitivity_plot}.  Note that these
rings with inclusive $\tau$ decays are much narrower
than the width of the contours or equivalently the
marginalized sensitivity on $\kappa_\tau$ after integrating
out the CP phase $\Delta$ from the original two-dimensional
distributions. The discrepancy comes from the fact that
the $\tau \rightarrow \pi \nu_\tau$ and
$\tau \rightarrow \rho \nu_\tau$ channels contribute
only a very small fraction ($\sim 13\%$) of the
inclusive decay events. The strength measurement can
provide very important complementary info and reduce
the parameter space to be explored.

\begin{figure}[t]
\centering
\includegraphics[scale = 0.35]{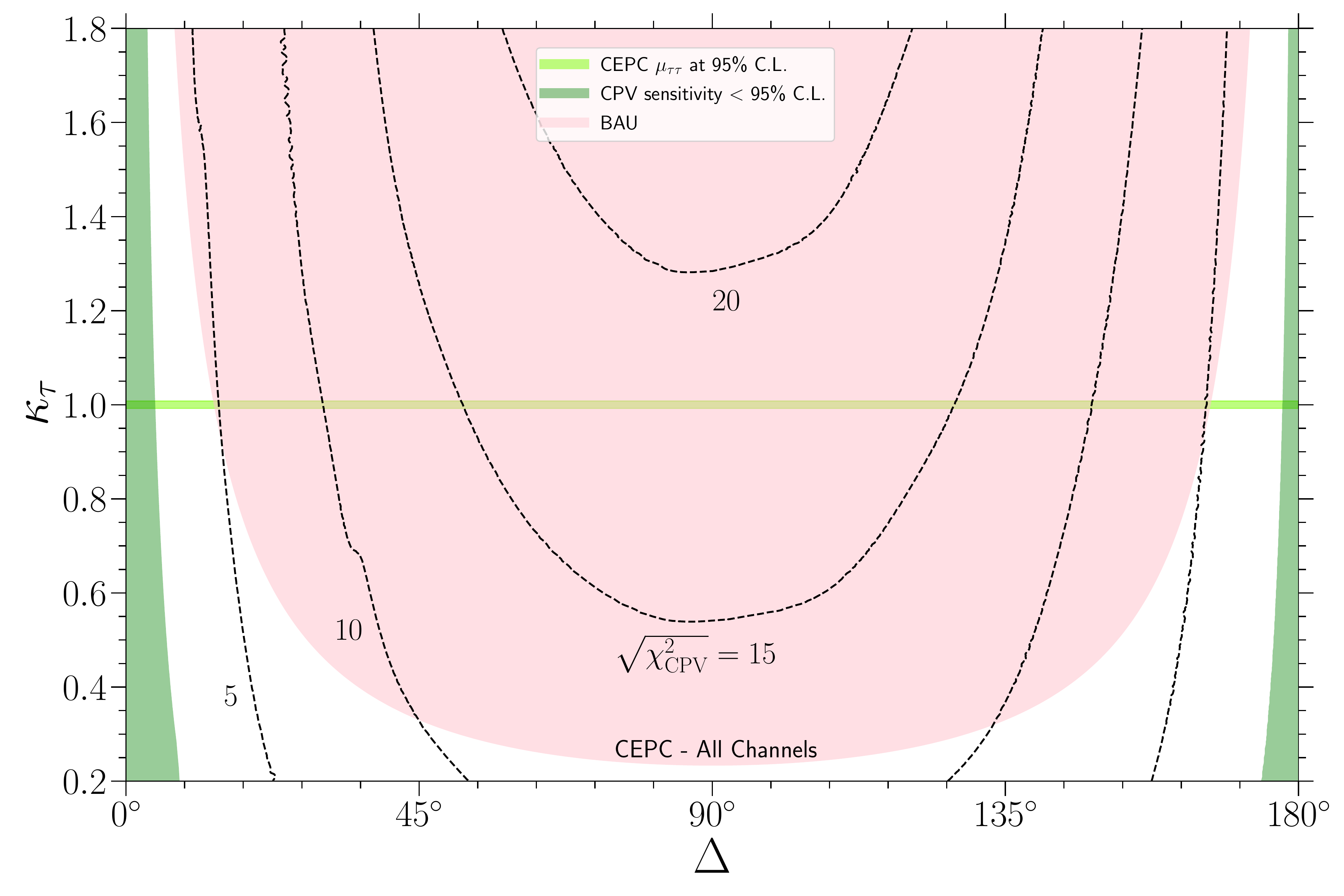}
\caption{The CP discovery capability of the CEPC as a
function of the $\Delta$ and $\kappa_\tau$ true values.
The black dashed lines represents several typical values of
the significance, $\sqrt{ \chi^2_{\rm CPV}} = 5,10,15,20$.
The green region represents the space parameter
where the sensitivity is below 95\% C.L. The pink region
represents the parameter space that can explain the BAU in the lepton flavored EWBG scenario \cite{Guo:2016ixx}.
}
    \label{fig:CPV_plot_varying_absY}
\end{figure}

Instead of assuming the SM values $\kappa^{\rm true}_\tau=1$ and $\Delta^{\rm true}=0^\circ$, it is interesting to ask the whether the lepton flavored EWBG scenario can explain the BAU and at the same time produce a signal that is distinguishable from the SM. 
To address this question, we show 
in the right panel of \gfig{fig:sensitivity_plot}
the similar contours around
$\Delta^{\rm true} = \pm 13.3^\circ$ and
$\kappa^{\rm true}_\tau = 1$ that is at the boundary of
the BAU region. 
Under this assumption, the CEPC and FCC could establish the presence of CPV in the $\tau$ Yukawa interaction with $3\sigma$ significance, while for the ILC the significance would be somewhat weaker.

It is also interesting to investigate the behavior of the CP violation sensitivity when one varies the assumed true values of $\kappa_\tau$.
This can be observed from 
\gfig{fig:CPV_plot_varying_absY} where we show the sensitivity as a function of the
CP phase $\Delta$ and the coupling strength $\kappa_\tau$.
The dashed gray lines
give several typical sensitivities
$\sqrt{\chi^2_{\rm CPV}} = 5, 10, 15, 20$. Note that the
dashed gray lines expand with larger $\tau$ Yukawa coupling
due to event number enhancement. This is especially
significant for small $\kappa_\tau$ while for large values of
$\kappa_\tau$
the CP sensitivity does not change substantially.
The BAU-compatible region has a lower limit at $\kappa_\tau \approx 0.25$ due to the
lower limit on $\kappa_\tau \sin \Delta$ according
to \gfig{fig:sensitivity_plot} and most of
the BAU-compatible region falls inside the
$\sqrt{ \chi^2_{\rm CPV}} = 5$ curve, corresponding
to $5\,\sigma$ discovery.

\section{Conclusions}
\label{sec:conclude}

Explaining the origin of the baryon asymmetry of the Universe is a key open problem at the interface of particle and nuclear physics with cosmology. An essential ingredient in the explanation is the presence of BSM CP violation. In the electroweak baryogenesis scenario, the relevant CPV interactions would have generated the BAU during the era of EWSB. The corresponding mass scale makes these interactions in principle experimentally accessible. While null results for permanent EDM searches place strong constraints on new flavor diagonal, electroweak scale CPV interactions, flavor changing CPV effects are significantly less restricted. Lepton flavored EWBG draws on this possibility, with interesting implications for CPV in the tau-lepton Yukawa sector.

In this work, we have shown how measurements of CPV observable in Higgs di-tau decays at prospective future lepton colliders could test this possibility, with significant discovery potential if it is realized in nature. After making a detailed comparison of the four
differential distributions of the neutrino azimuth
angle $\delta \phi_\nu$, polarimeter $\delta \phi_r$,
acoplanarity $\phi^*$, and the $\Theta$ variable for
the first time as well as various detector responses,
we explore the prospects of CP measurement in the
$\tau$ Yukawa coupling at future lepton colliders.
With $(5.6, 5, 2) \,\mbox{ab}^{-1}$ luminosity,
the $1\,\sigma$ uncertainty can reach
$2.9^\circ, 3.2^\circ, 3.8^\circ$ at the CEPC, FCC-ee,
and ILC, respectively. This allows the possibility
of distinguishing the attainable EWBG from the
CP conserving case with $3\,\sigma$ sensitivity.
The future lepton colliders are not just precision
machines for detailing our understanding of the
Higgs boson, but can also make $\mathcal O(1)$
measurement of the possible new physics beyond the
SM.

\section*{Acknowledgements}
SFG is sponsored by the Double First Class start-up fund (WF220442604) provided by Tsung-Dao Lee Institute, Shanghai Jiao Tong University and the Shanghai Pujiang Program (20PJ1407800).
SFG is also grateful to Kai Ma for sharing his PhD thesis
with derivations on the differential distribution of the
$\tau \rightarrow \pi \nu_\tau$ decay as well as
Manqi Ruan, Xin Chen, and Dan Yu for useful discussions.
GL would like to thank Shou-hua Zhu for helpful discussions. MJRM was supported in part under National Natural Science Foundation of China grant number 19Z103010239. GL and MJRM were supported in part under U.S. Department of Energy contract number DE-SC0011095.

\addcontentsline{toc}{section}{{References}}

\end{document}